\documentclass[draft,tightenlines,nofootinbib,preprint,aps,eqsecnum,amsmath,amssymb]{revtex4}

\newcommand{\beq}{\begin{equation}}
\newcommand{\eeq}{\end{equation}}
\newcommand{\bea}{\begin{eqnarray}}
\newcommand{\eea}{\end{eqnarray}}
\newcommand{\cir}{{\buildrel \circ \over =}}
\newcommand{\sgn}{\epsilon}

\begin{document}

\title{Hamiltonian Expression of Curvature Tensors in the York Canonical Basis:
II) The Weyl Tensor, Weyl Scalars, the Weyl Eigenvalues and the Problem of the Observables
of the Gravitational Field}

\medskip

\author{Luca Lusanna}

\affiliation{ Sezione INFN di Firenze\\ Polo Scientifico\\ Via Sansone 1\\
50019 Sesto Fiorentino (FI), Italy\\ E-mail: lusanna@fi.infn.it}

\author{Mattia Villani}

\affiliation{Dipartimento di Fisica, Universita' di Firenze\\ Polo Scientifico, Via Sansone 1,
50019 Sesto Fiorentino (FI), Italy\\ E-mail: villani@fi.infn.it}

\today

\bigskip

\begin{abstract}

We find the Hamiltonian expression in the York basis of canonical ADM tetrad gravity of the 4-Weyl tensor
of the asymptotically Minkowskian space-time. Like for the 4-Riemann tensor we find a radar tensor
(whose components are 4-scalars due to the use of radar 4-coordinates), which coincides with the 4-Weyl tensor on-shell on the solutions of Einstein's equations. Then, by using the Hamiltonian null tetrads, we find the Hamiltonian expression of the Weyl scalars of the Newman-Perose approach and of the four eigenvalues of the 4-Weyl tensor.

After having introduced the Dirac observables of canonical gravity, whose determination requires the solution of the super-Hamiltonian and super-momentum constraints, we discuss the connection of the Dirac observables with the notion of 4-scalar Bergmann observables. Due to the use of radar 4-coordinates  these two types of observables coincide in our formulation of canonical ADM tetrad gravity. However, contrary to Bergmann proposal, the Weyl eigenvalues are shown not to be Bergmann observables, so that their relevance is only in their use (first suggested by Bergmann and Komar) for giving a physical identification as point-events of the mathematical points of the space-time 4-manifold.

Finally we give the expression of the Weyl scalars in the Hamiltonian Post-Minkowskian linearization of canonical ADM tetrad gravity in the family of (non-harmonic) 3-orthogonal Schwinger time gauges.

\end{abstract}

\maketitle

\vfill\eject

\section{Introduction.}

The results of the first paper \cite{1}, which will be denoted as I in what follows with its equations denoted as
I-(..), will be used in this second paper to get the Hamiltonian 4-Weyl radar tensor (equal to the 4-Weyl tensor on-shell) and the Hamiltonian expression of the Weyl scalars of the Newman-Penrose formalism \cite{2,3} in the York basis of the formulation of canonical ADM tetrad gravity developed in Refs. \cite{4,5,6,7}. This will allow us to get the Hamiltonian expression of the four Weyl eigenvalues (4-scalars independent from the choice of the null tetrads).

\bigskip

Then we will show that if one would be able to solve the super-Hamiltonian and super-momentum constraints then it
would be possible to find a Shanmugadashan canonical transformation from the York basis to a final canonical basis adapted to all the 14 first-class constraints of canonical ADM tetrad gravity. This would allow to find the final
{\it Dirac observables} (DO) of the gravitational field as two conjugate pairs of canonical variables invariant under all the Hamiltonian gauge transformations. Like the tidal variables of the York basis, these DO's would  be  3-scalars of the instantaneous 3-spaces of the 3+1 splitting of the space-time and  4-scalars of the space-time due to our use
of radar 4-coordinates. Therefore these DO's would  be invariant under the group of passive 4-diffeomorphisms of the space-time, which is the gauge group of the generally covariant Lagrangian description of the gravitational field.
Moreover, in this final canonical basis the 4-scalar primary and secondary gauge variables would be completely separated from the DO's and the fixation of a gauge (at least the primary gauge fixings) could be made independently from them.
\medskip

In Refs.\cite{8} Bergmann proposed the notion of the so-called {\it Bergmann observables} (BO) as 4-scalar quantities
of the space-time uniquely predictable from a given set of initial data for Einstein's equations. In Ref.\cite{9} there is a full discussion on the self-consistency of this notion of BO ending with the enunciation of the conjecture that there should exist some canonical basis adapted to all the first class constraints whose DO's are also BO's. The finding of final Shanmugadhasan canonical transformation adapted to all the first class constraints in our framework using radar 4-coordinates would be the confirmation  of the validity of the conjecture.

\medskip

Instead the Bergmann-Komar proposal \cite{8} that a possible set of BO's for the gravitational field could be given by the four eigenvalues of the 4-Weyl tensor does not seem to be correct: in the York canonical basis these eigenvalues are complicated 4-scalar functions of both the tidal variables and  the inertial gauge variables.Therefore they are determined not only by the initial conditions for the tidal variables but also on the choice of the gauge. To arrive at a final conclusion one would need the expression of the Weyl eigenvalues in the final Shanmugadhasan canonical basis adapted to all the first class constraints.

\medskip

We will show that, as proposed in   Refs.\cite{10} following suggestions of Bergmann and Komar  in Refs.\cite{8},
the real utility of the Weyl eigenvalues is the definition of a special set of {\it intrinsic radar 4-coordinates}
allowing to give a physical (i.e. in terms of the gravitational field) individuation as point-events of the mathematical points of the space-time.

\bigskip

Finally we will give the Hamiltonian Post-Minkowskian (HPM) linearization (see Ref.\cite{6}) of the Weyl eigenvalues in the family of (non-harmonic) 3-orthogonal Schwinger time gauges used in Refs.\cite{5,6,7}. In these gauges only the Hamilton equations for the tidal variables are hyperbolic partial differential equations (PDE); all the constraints and the equations determining the lapse and shift functions are elliptic equations inside the 3-spaces of the 3+1 splitting of the asymptotically Minkowskian space-time.

\bigskip

In Section II we evaluate the Hamiltonian expression of the 4-Weyl tensor, of the Weyl scalars and of the Weyl eigenvalues. Then in Section III we give the Hamiltonian expression of the electric and magnetic components of the 4-Weyl tensor with respect to the congruence  of Eulerian observers associated with the 3+1 splitting of the space-time. In Appendix A there is the Hamiltonian expression of the Bel-Robinson tensor and of the second order invariants of the 4-Riemann and 4-Weyl tensors.

In Section IV we discuss the strategy to find the DO's of the gravitational field in arbitrary gauges by means of Shanmugadhasan canonical transformations and what can be said about the BO's.

In Section V we give the Hamiltonian expression of the Weyl scalars and of the Weyl eigenvalues in the HPM linearization of canonical ADM tetrad gravity in the family of 3-orthogonal Schwinger time gauges, whose theory is reviewed in Appendix B.

In the Conclusions there are some comments on the open problems to arrive at canonical quantization of gravity
in the framework presented in these two papers.

\section{The Weyl Tensor, the Weyl Scalars and the Weyl Eigenvalues.}

By using the Hamiltonian expression of the 4-Riemann and 4-Ricci tensors found in paper I and the Hamiltonian null tetrads there defined, we will find the Hamiltonian expression of the 4-Weyl tensor, of the Weyl scalars and of the Weyl eigenvalues in this Section.

\subsection{The 4-Weyl Tensor}

The traceless 4-Weyl tensor, with with only 10 independent components (and not 20 like the 4-Riemann tensor) due to its
symmetries, is

\medskip

\bea
 {}^4C_{ABCD} &=& {}^4R_{ABCD} - {1\over 2}\, ({}^4g_{AC}\, {}^4R_{BD} +
{}^4g_{BD}\, {}^4R_{AC} - {}^4g_{AD}\, {}^4R_{BC} - {}^4g_{BC}\, {}^4R_{AD}) +\nonumber \\
 &+&{1\over 6}\, ({}^4g_{AC}\, {}^4g_{BD} - {}^4g_{AD}\, {}^4g_{BC})\, {}^4R,\nonumber \\
 &&{}\nonumber\\
 &&{}^4C_{ABCD} = {}^4C_{CDAB} = - {}^4C_{BACD} = - {}^4C_{ABDC},\nonumber  \\
 &&{}^4C_{ABCD} + {}^4C_{ADBC} + {}^4C_{ACDB} = 0,\qquad
 {}^4g^{AC}\, {}^4C_{ABCD} = 0.\nonumber \\
 &&{}
 \label{2.1}
 \eea

\medskip

Eqs. I-(3.5) and I-(3.6) imply the following Hamiltonian expression for the components of the 4-Weyl radar tensor
($E_{AB} \cir 0$ are Einstein's equations I-(2.16))

\medskip

\begin{eqnarray*}
 {}^4C_{\tau r\tau s} &=& {}^4R_{\tau r\tau s} - {1\over 2}\,
 \Big({}^4g_{\tau\tau}\, {}^4R_{rs} + {}^4g_{rs}\, {}^4R_{\tau\tau} -\nonumber \\
 &-& {}^4g_{\tau s}\, {}^4R_{\tau r} - {}^4g_{\tau r}\, {}^4R_{\tau s}\Big) + {1\over
 6}\, \Big({}^4g_{\tau\tau}\, {}^4g_{rs} - {}^4g_{\tau r}\, {}^4g_{\tau s}\Big)\, {}^4R
 =\nonumber \\
 &&{}\nonumber \\
 &=& {\bar {\cal C}}_{\tau r\tau s} + {{\epsilon}\over 2}\, \Big([3\, (1 + n)^2 - {\bar n}_{(a)}\, {\bar n}_{(a)}]\,
 E_{rs} +\nonumber \\
 &+& {\tilde \phi}^{2/3}\, [Q^2_a\, V_{ra}\, V_{sa}\, E_{\tau\tau} - Q_a\, {\bar n}_{(a)}\, (V_{ra}\,
 E_{\tau s} + V_{sa}\, E_{\tau r})] \Big) -\nonumber \\
 &-& {1\over 6}\, {\tilde \phi}^{2/3}\, Q_a\, Q_b\, V_{ra}\, V_{sb}\, \Big([(1 + n)^2 - {\bar n}_{(a)}\, {\bar n}_{(a)}]\, \delta_{(a)(b)} + {\bar n}_{(a)}\, {\bar n}_{(b)}\Big)\, {}^4g^{AB}\, E_{AB} =\nonumber \\
 {}&&\nonumber \\
 &=& {\bar {\cal C}}_{\tau r\tau s} + {{\sgn}\over 4}\, {\tilde \phi}^{2/3}\,
 Q_a\, Q_b\, V_{ra}\, V_{sb}\nonumber \\
 &&\Big( {2\over 3}\, \sgn\, {\cal E}\, \delta_{(a)(b)} + 2N\, ({\cal E}_{00}
 + {\cal E}_{22})\, {\hat {\bar n}}_{(a)}\, {\hat {\bar n}}_{(b)}
 +\nonumber \\
 &+& \Big[(1 + n + \sqrt{\sum_c\, {\bar n}^2_{(c)}}\,\,)^2\, {\cal E}_{00} +
 (1 + n - \sqrt{\sum_c\, {\bar n}^2_{(c)}}\,\,)^2\, {\cal E}_{22} + 4(1 + n)^2\, {\cal E}_{11} +
 \nonumber \\
 &+& {{\sgn}\over 6}\, {\cal E}\, \sum_c\, {\bar n}^2_{(c)}\Big]\, \Big[{\hat {\bar
 \epsilon}}_{(+)(a)}\, {\hat {\bar \epsilon}}_{(-)(b)} +
 {\hat {\bar \epsilon}}_{(-)(a)}\, {\hat {\bar \epsilon}}_{(+)(b)}\Big] +\nonumber \\
 &+& \sqrt{2}\, (1 + n)\, \Big[(1 + n - \sqrt{\sum_c\, {\bar n}^2_{(c)}}\,\,)\, {\cal E}_{01} +
 (1 + n + \sqrt{\sum_c\, {\bar n}^2_{(c)}}\,\,)\, {\cal E}_{12}\Big]\, \Big[{\hat {\bar n}}_{(a)}\,
 {\hat {\bar \epsilon}}_{(-)(b)} + {\hat {\bar n}}_{(b)}\, {\hat {\bar
 \epsilon}}_{(-)(a)}\Big] +\nonumber \\
 &+& \sqrt{2}\, (1 + n)\, \Big[(1 + n - \sqrt{\sum_c\, {\bar n}^2_{(c)}}\,\,)\,
{\cal E}_{01}^* + (1 + n + \sqrt{\sum_c\, {\bar n}^2_{(c)}}\,\,)\,
{\cal E}_{12}^*\Big]\, \Big[{\hat {\bar n}}_{(a)}\, {\hat {\bar
\epsilon}}_{(+)(b)} + {\hat {\bar n}}_{(b)}\, {\hat {\bar
\epsilon}}_{(+)(a)}\Big] \Big),
  \end{eqnarray*}

 \begin{eqnarray*}
 {}^4C_{\tau ruv} &=& {}^4R_{\tau ruv} - {1\over 2}\,
 \Big({}^4g_{\tau u}\, {}^4R_{rv} + {}^4g_{rv}\, {}^4R_{\tau u} -\nonumber \\
 &-& {}^4g_{\tau v}\, {}^4R_{ru} - {}^4g_{ru}\, {}^4R_{\tau v}\Big) + {1\over
 6}\, \Big({}^4g_{\tau u}\, {}^4g_{rv} - {}^4g_{\tau v}\, {}^4g_{ru}\Big)\, {}^4R
=\nonumber \\
 &=& {\bar {\cal C}}_{\tau ruv} - {{\epsilon}\over 2}\, {\tilde \phi}^{1/3}\, Q_a\,
 \Big({\bar n}_{(a)}\, (V_{va}\, E_{ru} - V_{ua}\, E_{rv}) +\nonumber \\
 &+& {\tilde \phi}^{1/3}\, Q_a\, V_{ra}\, (V_{ua}\, E_{\tau v} - V_{va}\, E_{\tau u})\Big) +\nonumber \\
 &+& {1\over 6}\, \tilde \phi\, Q_a\, Q_b^2\, {\bar n}_{(a)}\, V_{rb}\, (V_{ua}\, V_{vb} - V_{va}\, V_{ub})\, {}^4g^{AB}\, E_{AB} =\nonumber \\
 \end{eqnarray*}

\begin{eqnarray*}
 &=& {\bar {\cal C}}_{\tau ruv} - {{\sgn}\over 4}\, \tilde \phi\,
 Q_a\, Q_b\, Q_c\, V_{ra}\, V_{ub}\, V_{vc}\nonumber \\
 &&\Big( \Big[(1 + n + \sqrt{\sum_c\, {\bar n}^2_{(c)}}\,\,)\, {\cal E}_{00} -
 (1 + n - \sqrt{\sum_c\, {\bar n}^2_{(c)}}\,\,)\, {\cal E}_{22} -\nonumber \\
 &-& 2\, \sqrt{\sum_c\, {\bar n}^2_{(c)}}\, ({\cal E}_{11} - {{\sgn}\over 3}\, {\cal E} )\Big]\,
 (\delta_{(a)(b)}\, {\hat {\bar n}}_{(c)} -
 \delta_{(a)(c)}\, {\hat {\bar n}}_{(b)}) +\nonumber \\
 \end{eqnarray*}

\begin{eqnarray*}
 &+& 2 {\cal E}_{11}\, \sqrt{\sum_c\, {\bar n}^2_{(c)}}\,
 \Big[ {\hat {\bar \epsilon}}_{(+)(a)}\, ({\hat {\bar \epsilon}}_{(-)(b)}\,
 {\hat {\bar n}}_{(c)} - {\hat {\bar \epsilon}}_{(c)}\, {\hat {\bar n}}_{(b)}) +
 {\hat {\bar \epsilon}}_{(-)(a)}\, ({\hat {\bar \epsilon}}_{(+)(b)}\,
 {\hat {\bar n}}_{(c)} - {\hat {\bar \epsilon}}_{(+)(c)}\, {\hat {\bar n}}_{(b)})\Big]
 +\nonumber \\
 &+& \sqrt{2}\, \Big[(1 + n + \sqrt{\sum_c\, {\bar n}^2_{(c)}}\,\,)\, {\cal E}_{01} -
 (1 + n - \sqrt{\sum_c\, {\bar n}^2_{(c)}}\,\,)\, {\cal E}_{12}\Big]\, (\delta_{(a)(b)}\, {\hat {\bar
 \epsilon}}_{(-)(c)} - \delta_{(a)(c)}\, {\hat {\bar
 \epsilon}}_{(-)(b)}) +\nonumber \\
 &+& \sqrt{2}\, \Big[(1 + n + \sqrt{\sum_c\, {\bar n}^2_{(c)}}\,\,)\, {\cal E}_{01}^* -
 (1 + n - \sqrt{\sum_c\, {\bar n}^2_{(c)}}\,\,)\, {\cal E}_{12}^*\Big]\, (\delta_{(a)(b)}\, {\hat {\bar
 \epsilon}}_{(+)(c)} - \delta_{(a)(c)}\, {\hat {\bar
 \epsilon}}_{(+)(b)}) +\nonumber \\
 &+& \sqrt{2}\, \sqrt{\sum_c\, {\bar n}^2_{(c)}}\, \Big[({\cal E}_{01} -
 {\cal E}_{12})\, {\hat {\bar n}}_{(a)}\, ({\hat {\bar \epsilon}}_{(-)(b)}\,
 {\hat {\bar n}}_{(c)} - {\hat {\bar \epsilon}}_{(-)(c)}\, {\hat {\bar n}}_{(b)}) +
 \nonumber \\
 &+& ({\cal E}_{01}^* - {\cal E}_{12}^*)\, {\hat {\bar n}}_{(a)}\, ({\hat {\bar \epsilon}}_{(+)(b)}\,
 {\hat {\bar n}}_{(c)} - {\hat {\bar \epsilon}}_{(+)(c)}\, {\hat {\bar n}}_{(b)})
 \Big] -\nonumber \\
 &-& 2\, \sqrt{\sum_c\, {\bar n}^2_{(c)}}\, \Big[{\cal E}_{02}\, {\hat {\bar
 \epsilon}}_{(-)(a)}\, ({\hat {\bar n}}_{(b)}\, {\hat {\bar \epsilon}}_{(-)(c)}
 - {\hat {\bar n}}_{(c)}\, {\hat {\bar \epsilon}}_{(-)(b)}) + {\cal E}_{02}^*\,
 {\hat {\bar \epsilon}}_{(+)(a)}\, ({\hat {\bar n}}_{(b)}\, {\hat {\bar \epsilon}}_{(+)(c)} -
 {\hat {\bar n}}_{(c)}\, {\hat {\bar \epsilon}}_{(+)(b)})\Big]\Big),
  \end{eqnarray*}

 \begin{eqnarray*}s
 {}^4C_{rsuv} &=& {}^4R_{rsuv} - {1\over 2}\,
 \Big({}^4g_{ru}\, {}^4R_{sv} + {}^4g_{sv}\, {}^4R_{ru} -\nonumber \\
 &-& {}^4g_{rv}\, {}^4R_{su} - {}^4g_{su}\, {}^4R_{rv}\Big) + {1\over
 6}\, \Big({}^4g_{ru}\, {}^4g_{sv} - {}^4g_{rv}\, {}^4g_{su}\Big)\, {}^4R
=\nonumber \\
 &=& {\bar {\cal C}}_{rsuv} + {{\epsilon}\over 2}\, {\tilde \phi}^{2/3}\, Q_a^2\, \Big(V_{ra}\, (V_{ua}\,
 E_{sv} - V_{va}\, E_{su})  + V_{sa}\,(V_{va}\, E_{ru} - V_{ua}\, E_{rv}) \Big) +\nonumber \\
 &+& {1\over 6}\, {\tilde \phi}^{4/3}\, Q_a^2\, Q_b^2\, V_{ra}\, V_{sb}\, (V_{ua}\, V_{vb} - V_{va}\, V_{ub})\,
 {}4g^{AB}\, E_{AB} =\nonumber \\
 {}&&\nonumber \\
 &=&  {\bar {\cal C}}_{rsuv} + {{\sgn}\over 4}\, {\tilde \phi}^{4/3}\, Q_a\, Q_b\, Q_c\, Q_d\,
 V_{ra}\, V_{sb}\, V_{uc}\, V_{vd}\nonumber \\
 &&\Big( - {{\sgn}\over 3}\, {\cal E}\, (\delta_{(a)(c)}\, \delta_{(b)(d)} -
 \delta_{(a)(d)}\, \delta_{(b)(c)}) -\nonumber \\
 &-& 2\, {\cal E}_{11}\, \Big[\delta_{(a)(c)}\, ({\hat {\bar n}}_{(b)}\,
 {\hat {\bar n}}_{(d)} - {\hat {\bar \epsilon}}_{(+)(b)}\, {\hat {\bar \epsilon}}_{(-)(d)}
 - {\hat {\bar \epsilon}}_{(-)(b)}\, {\hat {\bar \epsilon}}_{(+)(d)}) +\nonumber \\
 &+& \delta_{(b)(d)}\, ({\hat {\bar n}}_{(a)}\,
 {\hat {\bar n}}_{(c)} - {\hat {\bar \epsilon}}_{(+)(a)}\, {\hat {\bar \epsilon}}_{(-)(c)}
 - {\hat {\bar \epsilon}}_{(-)(a)}\, {\hat {\bar \epsilon}}_{(+)(c)}) -\nonumber \\
 &-& \delta_{(a)(d)}\, ({\hat {\bar n}}_{(b)}\,
 {\hat {\bar n}}_{(c)} - {\hat {\bar \epsilon}}_{(+)(b)}\, {\hat {\bar \epsilon}}_{(-)(c)}
 - {\hat {\bar \epsilon}}_{(-)(b)}\, {\hat {\bar \epsilon}}_{(+)(c)}) -\nonumber \\
 &-& \delta_{(b)(c)}\, ({\hat {\bar n}}_{(a)}\,
 {\hat {\bar n}}_{(d)} - {\hat {\bar \epsilon}}_{(+)(a)}\, {\hat {\bar \epsilon}}_{(-)(d)}
 - {\hat {\bar \epsilon}}_{(-)(a)}\, {\hat {\bar \epsilon}}_{(+)(d)}) \Big]
 +\nonumber \\
 &+& ({\cal E}_{00} + {\cal E}_{22})\, \Big[\delta_{(a)(c)}\, {\hat {\bar n}}_{(b)}\,
 {\hat {\bar n}}_{(d)} + \delta_{(b)(d)}\, {\hat {\bar n}}_{(a)}\, {\hat {\bar n}}_{(c)}
 - \delta_{(a)(d)}\, {\hat {\bar n}}_{(b)}\, {\hat {\bar n}}_{(c)} -
 \delta_{(b)(c)}\, {\hat {\bar n}}_{(a)}\, {\hat {\bar n}}_{(d)}\Big] +\nonumber \\
 &+& 2\, {\cal E}_{02}\, \Big[\delta_{(a)(c)}\, {\hat {\bar \epsilon}}_{(-)(b)}\,
 {\hat {\bar \epsilon}}_{(-)(d)} + \delta_{(b)(d)}\, {\hat {\bar \epsilon}}_{(-)(a)}\,
 {\hat {\bar \epsilon}}_{(-)(c)} -\nonumber \\
 &-& \delta_{(a)(d)}\, {\hat {\bar \epsilon}}_{(-)(b)}\,
 {\hat {\bar \epsilon}}_{(-)(c)} - \delta_{(b)(c)}\, {\hat {\bar \epsilon}}_{(-)(a)}\,
 {\hat {\bar \epsilon}}_{(-)(d)} \Big] +\nonumber \\
 &+& 2\, {\cal E}_{02}^*\, \Big[\delta_{(a)(c)}\, {\hat {\bar \epsilon}}_{(+)(b)}\,
 {\hat {\bar \epsilon}}_{(+)(d)} + \delta_{(b)(d)}\, {\hat {\bar \epsilon}}_{(+)(a)}\,
 {\hat {\bar \epsilon}}_{(+)(c)} -\nonumber \\
 &-& \delta_{(a)(d)}\, {\hat {\bar \epsilon}}_{(+)(b)}\,
 {\hat {\bar \epsilon}}_{(+)(c)} - \delta_{(b)(c)}\, {\hat {\bar \epsilon}}_{(+)(a)}\,
 {\hat {\bar \epsilon}}_{(+)(d)} \Big] +\nonumber \\
 \end{eqnarray*}

\bea
 &+& \sqrt{2}\, ({\cal E}_{01} - {\cal E}_{12})\, \Big[\delta_{(a)(c)}\,
 ({\hat {\bar n}}_{(b)}\, {\hat {\bar \epsilon}}_{(-)(d)} + {\hat {\bar n}}_{(d)}\,
 {\hat {\bar \epsilon}}_{(-)(b)}) + \delta_{(b)(d)}\, ({\hat {\bar n}}_{(a)}\,
 {\hat {\bar \epsilon}}_{(-)(c)} + {\hat {\bar n}}_{(c)}\,
 {\hat {\bar \epsilon}}_{(-)(a)}) -\nonumber \\
 &-& \delta_{(a)(d)}\, ({\hat {\bar n}}_{(b)}\, {\hat {\bar \epsilon}}_{(-)(c)} +
 {\hat {\bar n}}_{(c)}\, {\hat {\bar \epsilon}}_{(-)(b)}) - \delta_{(b)(c)}\,
  ({\hat {\bar n}}_{(a)}\, {\hat {\bar \epsilon}}_{(-)(d)} + {\hat {\bar n}}_{(d)}\,
 {\hat {\bar \epsilon}}_{(-)(a)}) \Big] +\nonumber \\
 &+& \sqrt{2}\, ({\cal E}_{01}^* - {\cal E}_{12}^*)\, \Big[\delta_{(a)(c)}\,
 ({\hat {\bar n}}_{(b)}\, {\hat {\bar \epsilon}}_{(+)(d)} + {\hat {\bar n}}_{(d)}\,
 {\hat {\bar \epsilon}}_{(+)(b)}) + \delta_{(b)(d)}\, ({\hat {\bar n}}_{(a)}\,
 {\hat {\bar \epsilon}}_{(+)(c)} + {\hat {\bar n}}_{(c)}\,
 {\hat {\bar \epsilon}}_{(+)(a)}) -\nonumber \\
 &-& \delta_{(a)(d)}\, ({\hat {\bar n}}_{(b)}\, {\hat {\bar \epsilon}}_{(+)(c)} +
 {\hat {\bar n}}_{(c)}\, {\hat {\bar \epsilon}}_{(+)(b)}) - \delta_{(b)(c)}\,
  ({\hat {\bar n}}_{(a)}\, {\hat {\bar \epsilon}}_{(+)(d)} + {\hat {\bar n}}_{(d)}\,
 {\hat {\bar \epsilon}}_{(+)(a)}) \Big] \Big).
  \label{2.2}
\end{eqnarray}

\noindent Then Eqs. I-(4.7) and I-(4.8) allow us to rewrite the 4-Weyl tensor in terms of the components of the energy-momentum tensor of matter and of Einstein's equations in the null tetrad basis I-(4.5). The properties of the Hamiltonian functions ${\bar W}_{....}$ given in Eq. I-(3.6) imply ${}^4g^{AC}\, {}^4C_{ABCD} = 0$. Eqs.I-(4.9) and I-(4.4) have been used for the expression in the basis of null tetrads.

 \bigskip

The previous equations define the following Hamiltonian radar tensor which coincides
with the 4-Weyl tensor on-shell on the solutions of Einstein's equations ( ${\bar W}_{....}$ are Hamiltonian functions defined in the equations I-(3.5) and I-(3.6); Eqs. I-(4.7) and I-(4.4) have been used for the expression in the basis of null tetrads)

\begin{eqnarray*}
 {\bar {\cal C}}_{ABCD} &=& {}^4C_{ABCD} - (terms\, proportional\, to\, Einstein's\,
 equations)\,\,\, \cir {}^4C_{ABCD},
 \end{eqnarray*}

 \begin{eqnarray*}
 {\bar {\cal C}}_{\tau r\tau s} &{\buildrel {def}\over =}& {\bar W}_{\tau r\tau s} + {{8\pi G}\over {c^3}}\, \Big[
 {{\sgn}\over 2}\,  \Big(((1 + n)^2 + {\bar n}_{(a)}\, {\bar n}_{(a)})\,
 {\hat T}_{rs} + {\tilde \phi}^{2/3}\, Q_a^2\, V_{ra}\, V_{sa}\,
 {\hat T}_{\tau\tau} -\nonumber \\
 &-& {\tilde \phi}^{1/3}\, Q_a\, (V_{ra}\, {\hat T}_{\tau s} +
 V_{sa}\, {\hat T}_{\tau r})\, {\bar n}_{(a)}\Big) -\nonumber \\
 &-& {1\over 6}\, {\tilde \phi}^{2/3}\,
 \Big(((1 + n)^2 - {\bar n}_{(c)}\, {\bar n}_{(c)})\, Q_a^2\,
 V_{ra}\, V_{sa} - Q_a\, Q_b\, V_{ra}\, V_{sb}\, {\bar n}_{(a)}\, {\bar
 n}_{(b)}\Big)\, T \Big] =\nonumber \\
 &&{}\nonumber \\
 &=& {\bar W}_{\tau r\tau s} + \sgn\, {{2 \pi G}\over {c^3}}\, {\tilde \phi}^{2/3}\,
 Q_a\, Q_b\, V_{ra}\, V_{sb}\nonumber \\
 &&\Big( \sgn\, {2\over 3}\, {\cal T}\, \delta_{(a)(b)} + 2N\, ({\cal T}_{00}
 + {\cal T}_{22})\, {\hat {\bar n}}_{(a)}\, {\hat {\bar n}}_{(b)}
 +\nonumber \\
 &+& \Big[(1 + n + \sqrt{\sum_c\, {\bar n}^2_{(c)}}\,\,)^2\, {\cal T}_{00} +
 (1 + n - \sqrt{\sum_c\, {\bar n}^2_{(c)}}\,\,)^2\, {\cal T}_{22} + 4(1 + n)^2\, \Phi_{11} +
 \nonumber \\
 &+& {{\sgn}\over 6}\, {\cal T}\, \sum_c\, {\bar n}^2_{(c)}\, \Big]\, \Big[{\hat {\bar
 \epsilon}}_{(+)(a)}\, {\hat {\bar \epsilon}}_{(-)(b)} +
 {\hat {\bar \epsilon}}_{(-)(a)}\, {\hat {\bar \epsilon}}_{(+)(b)}\Big] +\nonumber \\
 &+& \sqrt{2}\, (1 + n)\, \Big[(1 + n - \sqrt{\sum_c\, {\bar n}^2_{(c)}}\,\,)\, {\cal T}_{01} +
 (1 + n + \sqrt{\sum_c\, {\bar n}^2_{(c)}}\,\,)\, {\cal T}_{12}\Big]\, \Big[{\hat {\bar n}}_{(a)}\,
 {\hat {\bar \epsilon}}_{(-)(b)} + {\hat {\bar n}}_{(b)}\, {\hat {\bar
 \epsilon}}_{(-)(a)}\Big] +\nonumber \\
 &+& \sqrt{2}\, (1 + n)\, \Big[(1 + n - \sqrt{\sum_c\, {\bar n}^2_{(c)}}\,\,)\,
{\cal T}_{01}^* + (1 + n + \sqrt{\sum_c\, {\bar n}^2_{(c)}}\,\,)\,
{\cal T}_{12}^*\Big]\, \Big[{\hat {\bar n}}_{(a)}\, {\hat {\bar
\epsilon}}_{(+)(b)} + {\hat {\bar n}}_{(b)}\, {\hat {\bar
\epsilon}}_{(+)(a)}\Big] \Big),
  \end{eqnarray*}

 \begin{eqnarray*}
 {\bar {\cal C}}_{\tau ruv} &{\buildrel {def}\over =}&  {\bar W}_{\tau ruv} + {{8 \pi G}\over {c^3}}\, \Big[-
 {{\sgn}\over 2}\, \Big({\tilde \phi}^{1/3}\, Q_a\, (V_{va}\, {\hat
 T}_{ru} - V_{ua}\, {\hat T}_{rv})\, {\bar n}_{(a)}+\nonumber \\
 &+& {\tilde \phi}^{2/3}\, Q_a^2\, V_{ra}\, (V_{ua}\, {\hat T}_{\tau
 r} - V_{va}\, {\hat T}_{\tau u})\Big) + {1\over 6}\, \tilde \phi\,
 Q_a\, Q_b^2\, V_{rb}\, (V_{ua}\, V_{vb} - V_{va}\, V_{ub})\, {\bar
 n}_{(a)}\, T \Big] =\nonumber \\
  {}&& \nonumber \\
  &=& {\bar W}_{\tau ruv} - \sgn\, {{2 \pi G}\over {c^3}}\, \tilde \phi\,
 Q_a\, Q_b\, Q_c\, V_{ra}\, V_{ub}\, V_{vc}\nonumber \\
 &&\Big( \Big[(1 + n + \sqrt{\sum_c\, {\bar n}^2_{(c)}}\,\,)\, {\cal T}_{00} -
 (1 + n - \sqrt{\sum_c\, {\bar n}^2_{(c)}}\,\,)\, {\cal T}_{22} -\nonumber \\
 &-& 2\, \sqrt{\sum_c\, {\bar n}^2_{(c)}}\, ({\cal T}_{11} - {{\sgn}\over 3}\, {\cal T} )\Big]\,
 (\delta_{(a)(b)}\, {\hat {\bar n}}_{(c)} -
 \delta_{(a)(c)}\, {\hat {\bar n}}_{(b)}) +\nonumber \\
 &+& 2 {\cal T}_{11}\, \sqrt{\sum_c\, {\bar n}^2_{(c)}}\,
 \Big[ {\hat {\bar \epsilon}}_{(+)(a)}\, ({\hat {\bar \epsilon}}_{(-)(b)}\,
 {\hat {\bar n}}_{(c)} - {\hat {\bar \epsilon}}_{(c)}\, {\hat {\bar n}}_{(b)}) +
 {\hat {\bar \epsilon}}_{(-)(a)}\, ({\hat {\bar \epsilon}}_{(+)(b)}\,
 {\hat {\bar n}}_{(c)} - {\hat {\bar \epsilon}}_{(+)(c)}\, {\hat {\bar n}}_{(b)})\Big]
 +\nonumber \\
 &+& \sqrt{2}\, \Big[(1 + n + \sqrt{\sum_c\, {\bar n}^2_{(c)}}\,\,)\, {\cal T}_{01} -
 (1 + n - \sqrt{\sum_c\, {\bar n}^2_{(c)}}\,\,)\, {\cal T}_{12}\Big]\, (\delta_{(a)(b)}\, {\hat {\bar
 \epsilon}}_{(-)(c)} - \delta_{(a)(c)}\, {\hat {\bar
 \epsilon}}_{(-)(b)}) +\nonumber \\
 &+& \sqrt{2}\, \Big[(1 + n + \sqrt{\sum_c\, {\bar n}^2_{(c)}}\,\,)\, {\cal T}_{01}^* -
 (1 + n - \sqrt{\sum_c\, {\bar n}^2_{(c)}}\,\,)\, {\cal T}_{12}^*\Big]\, (\delta_{(a)(b)}\, {\hat {\bar
 \epsilon}}_{(+)(c)} - \delta_{(a)(c)}\, {\hat {\bar
 \epsilon}}_{(+)(b)}) +\nonumber \\
 &+& \sqrt{2}\, \sqrt{\sum_c\, {\bar n}^2_{(c)}}\, \Big[({\cal T}_{01} -
 {\cal T}_{12})\, {\hat {\bar n}}_{(a)}\, ({\hat {\bar \epsilon}}_{(-)(b)}\,
 {\hat {\bar n}}_{(c)} - {\hat {\bar \epsilon}}_{(-)(c)}\, {\hat {\bar n}}_{(b)}) +
 \nonumber \\
 &+& ({\cal T}_{01}^* - {\cal T}_{12}^*)\, {\hat {\bar n}}_{(a)}\, ({\hat {\bar \epsilon}}_{(+)(b)}\,
 {\hat {\bar n}}_{(c)} - {\hat {\bar \epsilon}}_{(+)(c)}\, {\hat {\bar n}}_{(b)})
 \Big] -\nonumber \\
 &-& 2\, \sqrt{\sum_c\, {\bar n}^2_{(c)}}\, \Big[{\cal T}_{02}\, {\hat {\bar
 \epsilon}}_{(-)(a)}\, ({\hat {\bar n}}_{(b)}\, {\hat {\bar \epsilon}}_{(-)(c)}
 - {\hat {\bar n}}_{(c)}\, {\hat {\bar \epsilon}}_{(-)(b)}) + {\cal T}_{02}^*\,
 {\hat {\bar \epsilon}}_{(+)(a)}\, ({\hat {\bar n}}_{(b)}\, {\hat {\bar \epsilon}}_{(+)(c)} -
 {\hat {\bar n}}_{(c)}\, {\hat {\bar \epsilon}}_{(+)(b)})\Big]\Big),
   \end{eqnarray*}

  \begin{eqnarray*}
  {\bar {\cal C}}_{rsuv} &{\buildrel {def}\over =}&
  {\bar W}_{rsuv} + {{8 \pi G}\over {c^3}}\, \Big[{{\sgn}\over
 2}\, {\tilde \phi}^{2/3}\, Q_a^2\, (V_{ra}\, V_{ua}\, {\hat T}_{sv}
 + V_{sa}\, V_{va}\, {\hat T}_{ru} - V_{ra}\, V_{va}\, {\hat T}_{su}
 - V_{sa}\, V_{ua}\, {\hat T}_{rv}) +\nonumber \\
 &-& {1\over 6}\, {\tilde \phi}^{4/3}\, Q_a^2\, Q_b^2\, V_{ra}\,
 V_{sb}\, (V_{ua}\, V_{vb} - V_{va}\, V_{ub})\, T \Big] =\nonumber \\
 {}&&\nonumber \\
 &=&  {\bar W}_{rsuv} + \sgn\, {{2 \pi G}\over {c^3}}\, {\tilde \phi}^{4/3}\, Q_a\, Q_b\, Q_c\, Q_d\,
 V_{ra}\, V_{sb}\, V_{uc}\, V_{vd}\nonumber \\
 &&\Big( - {{\sgn}\over 3}\, {\cal T}\, (\delta_{(a)(c)}\, \delta_{(b)(d)} -
 \delta_{(a)(d)}\, \delta_{(b)(c)}) -\nonumber \\
 &-& 2\, {\cal T}_{11}\, \Big[\delta_{(a)(c)}\, ({\hat {\bar n}}_{(b)}\,
 {\hat {\bar n}}_{(d)} - {\hat {\bar \epsilon}}_{(+)(b)}\, {\hat {\bar \epsilon}}_{(-)(d)}
 - {\hat {\bar \epsilon}}_{(-)(b)}\, {\hat {\bar \epsilon}}_{(+)(d)}) +\nonumber \\
 &+& \delta_{(b)(d)}\, ({\hat {\bar n}}_{(a)}\,
 {\hat {\bar n}}_{(c)} - {\hat {\bar \epsilon}}_{(+)(a)}\, {\hat {\bar \epsilon}}_{(-)(c)}
 - {\hat {\bar \epsilon}}_{(-)(a)}\, {\hat {\bar \epsilon}}_{(+)(c)}) -\nonumber \\
 &-& \delta_{(a)(d)}\, ({\hat {\bar n}}_{(b)}\,
 {\hat {\bar n}}_{(c)} - {\hat {\bar \epsilon}}_{(+)(b)}\, {\hat {\bar \epsilon}}_{(-)(c)}
 - {\hat {\bar \epsilon}}_{(-)(b)}\, {\hat {\bar \epsilon}}_{(+)(c)}) -\nonumber \\
 &-& \delta_{(b)(c)}\, ({\hat {\bar n}}_{(a)}\,
 {\hat {\bar n}}_{(d)} - {\hat {\bar \epsilon}}_{(+)(a)}\, {\hat {\bar \epsilon}}_{(-)(d)}
 - {\hat {\bar \epsilon}}_{(-)(a)}\, {\hat {\bar \epsilon}}_{(+)(d)}) \Big]
 +\nonumber \\
 &+& ({\cal T}_{00} + {\cal T}_{22})\, \Big[\delta_{(a)(c)}\, {\hat {\bar n}}_{(b)}\,
 {\hat {\bar n}}_{(d)} + \delta_{(b)(d)}\, {\hat {\bar n}}_{(a)}\, {\hat {\bar n}}_{(c)}
 - \delta_{(a)(d)}\, {\hat {\bar n}}_{(b)}\, {\hat {\bar n}}_{(c)} -
 \delta_{(b)(c)}\, {\hat {\bar n}}_{(a)}\, {\hat {\bar n}}_{(d)}\Big] +\nonumber \\
 \end{eqnarray*}

\bea
 &+& 2\, {\cal T}_{02}\, \Big[\delta_{(a)(c)}\, {\hat {\bar \epsilon}}_{(-)(b)}\,
 {\hat {\bar \epsilon}}_{(-)(d)} + \delta_{(b)(d)}\, {\hat {\bar \epsilon}}_{(-)(a)}\,
 {\hat {\bar \epsilon}}_{(-)(c)} -\nonumber \\
 &-& \delta_{(a)(d)}\, {\hat {\bar \epsilon}}_{(-)(b)}\,
 {\hat {\bar \epsilon}}_{(-)(c)} - \delta_{(b)(c)}\, {\hat {\bar \epsilon}}_{(-)(a)}\,
 {\hat {\bar \epsilon}}_{(-)(d)} \Big] +\nonumber \\
 &+& 2\, {\cal T}_{02}^*\, \Big[\delta_{(a)(c)}\, {\hat {\bar \epsilon}}_{(+)(b)}\,
 {\hat {\bar \epsilon}}_{(+)(d)} + \delta_{(b)(d)}\, {\hat {\bar \epsilon}}_{(+)(a)}\,
 {\hat {\bar \epsilon}}_{(+)(c)} -\nonumber \\
 &-& \delta_{(a)(d)}\, {\hat {\bar \epsilon}}_{(+)(b)}\,
 {\hat {\bar \epsilon}}_{(+)(c)} - \delta_{(b)(c)}\, {\hat {\bar \epsilon}}_{(+)(a)}\,
 {\hat {\bar \epsilon}}_{(+)(d)} \Big] +\nonumber \\
 &+& \sqrt{2}\, ({\cal T}_{01} - {\cal T}_{12})\, \Big[\delta_{(a)(c)}\,
 ({\hat {\bar n}}_{(b)}\, {\hat {\bar \epsilon}}_{(-)(d)} + {\hat {\bar n}}_{(d)}\,
 {\hat {\bar \epsilon}}_{(-)(b)}) + \delta_{(b)(d)}\, ({\hat {\bar n}}_{(a)}\,
 {\hat {\bar \epsilon}}_{(-)(c)} + {\hat {\bar n}}_{(c)}\,
 {\hat {\bar \epsilon}}_{(-)(a)}) -\nonumber \\
 &-& \delta_{(a)(d)}\, ({\hat {\bar n}}_{(b)}\, {\hat {\bar \epsilon}}_{(-)(c)} +
 {\hat {\bar n}}_{(c)}\, {\hat {\bar \epsilon}}_{(-)(b)}) - \delta_{(b)(c)}\,
  ({\hat {\bar n}}_{(a)}\, {\hat {\bar \epsilon}}_{(-)(d)} + {\hat {\bar n}}_{(d)}\,
 {\hat {\bar \epsilon}}_{(-)(a)}) \Big] +\nonumber \\
 &+& \sqrt{2}\, ({\cal T}_{01}^* - {\cal T}_{12}^*)\, \Big[\delta_{(a)(c)}\,
 ({\hat {\bar n}}_{(b)}\, {\hat {\bar \epsilon}}_{(+)(d)} + {\hat {\bar n}}_{(d)}\,
 {\hat {\bar \epsilon}}_{(+)(b)}) + \delta_{(b)(d)}\, ({\hat {\bar n}}_{(a)}\,
 {\hat {\bar \epsilon}}_{(+)(c)} + {\hat {\bar n}}_{(c)}\,
 {\hat {\bar \epsilon}}_{(+)(a)}) -\nonumber \\
 &-& \delta_{(a)(d)}\, ({\hat {\bar n}}_{(b)}\, {\hat {\bar \epsilon}}_{(+)(c)} +
 {\hat {\bar n}}_{(c)}\, {\hat {\bar \epsilon}}_{(+)(b)}) - \delta_{(b)(c)}\,
  ({\hat {\bar n}}_{(a)}\, {\hat {\bar \epsilon}}_{(+)(d)} + {\hat {\bar n}}_{(d)}\,
 {\hat {\bar \epsilon}}_{(+)(a)}) \Big] \Big).
 \label{2.3}
\eea

\medskip

In Appendix A there is the Hamiltonian expression of the Bel-Robinson tensor and of the second-order invariants of the
4-Weyl tensor.

\subsection{The Weyl Scalars}

By using the null tetrads I-(4.5) the 10 components of the 4-Weyl
tensor \footnote{In the Newman-Penrose formalism \cite{2} the 4-Weyl tensor
is replaced by a symmetric spinor $\psi_{(abcd)} = \eta^{(1)}_{(a}
\, \eta^{(2)}_b\, \eta^{(3)}_c\, \eta^{(4)}_{d)}$.} can be replaced by the following 5 complex Weyl scalars
(see for instance Ref.\cite{3}, pp.189-194)

\bea
 \Psi_0 &=& {}^4C_{ABCD}\, K^A\, M^B\, K^C\, M^D,\nonumber \\
 \Psi_1 &=& {}^4C_{ABCD}\, K^A\, L^B\, K^C\, M^D,\nonumber  \\
 \Psi_2 &=& {1\over 2}\, {}^4C_{ABCD}\, K^A\, L^B\, (K^C\, L^D - M^C\, M^{*D}),
 \nonumber \\
 \Psi_3 &=& {}^4C_{ABCD}\, L^A\, K^B\, L^C\, M^{*D},\nonumber \\
 \Psi_4 &=& {}^4C_{ABCD}\, L^A\, M^{*B}\, L^C\, M^{*D},
 \label{2.4}
 \eea

\noindent which will be the sum of terms  $\Psi^{(W)}_{\cal A}$
(${\cal A} = 0,1,..,4$), 4-scalars functions of the canonical
variables, plus terms vanishing with the Einstein's equations\medskip

\beq
 \Psi_{\cal A}  {\buildrel {def}\over =}\,\,
 \Psi^{(W)}_{\cal A} + \Psi^{(E)}_{\cal A},\qquad \Psi^{(E)}_{\cal A} \cir 0.
 \label{2.5}
 \eeq

\bigskip

By using Eqs.(\ref{2.2}) and (\ref{2.3}) the expression of the
Weyl scalars is (${}^3{\bar e}^r_{(a)} = {\tilde \phi}^{-1/3}\, Q_a^{-1}\, V_{ra}$)
\bigskip

\begin{eqnarray*}
 \Psi_0 &=&  {}^4C_{ABCD}\, K^A\, M^B\, K^C\, M^D =
\Psi_0^{(W)} + \Psi_0^{(E)},\nonumber \\
 &&{}\nonumber \\
 &&{} \nonumber \\
 \Psi_0^{(W)} &=& - {1\over 2}\, \Big({1\over {(1 + n)^2}}\,
 {\hat {\bar \epsilon}}_{(+)(a)}\, {\hat {\bar
 \epsilon}}_{(+)(b)}\, {\bar W}_{\tau r\tau s}\, {}^3{\bar e}^r_{(a)}\,
 {}^3{\bar e}^s_{(b)} +\nonumber \\
 &+& {2\over {1 + n}}\, \Big[1 - {{\sqrt{\sum_e\, {\bar n}^2_{(e)}}}\over {1 + n}}\Big]\,
 {\hat {\bar \epsilon}}_{(+)(a)}\, {\hat {\bar n}}_{(b)}\, {\hat
 {\bar \epsilon}}_{(+)(c)}\, {\bar W}_{\tau ruv}\, {}^3{\bar
 e}^r_{(a)}\,{}^3{\bar e}^u_{(b)}\, {}^3{\bar e}^v_{(c)}
 +\nonumber \\
 &+& \Big[1 - {{\sqrt{\sum_e\, {\bar n}^2_{(e)}}}\over {1 + n}}\Big]^2\,
 {\hat {\bar n}}_{(a)}\, {\hat {\bar \epsilon}}_{(+)(b)}\, {\hat
 {\bar n}}_{(c)}\, {\hat {\bar \epsilon}}_{(+)(d)}\, {\bar W}_{rsuv}\,
 {}^3{\bar e}^r_{(a)}\, {}^3{\bar e}^s_{(b)}\, {}^3{\bar
 e}^u_{(c)}\, {}^3{\bar e}^v_{(d)} \Big) +\nonumber \\
 &+& \sgn\, {{2\pi G}\over {c^3}}\, \Big(1 - {{\sum_c\,
 {\bar n}^2_{(c)}}\over {(1 + n)^2}}\Big)\, {\cal T}_{02},\nonumber \\
  &&{}\nonumber \\
 &&{}\nonumber \\
 \Psi_0^{(E)} &=& {{\sgn}\over 4}\, \Big(1 - {{\sum_c\,
 {\bar n}^2_{(c)}}\over {(1 + n)^2}}\Big)\, {\cal E}_{02} \cir 0,
 \end{eqnarray*}

 \medskip

 \begin{eqnarray*}
 \Psi_1 &=&  {}^4C_{ABCD}\, K^A\, L^B\, K^C\, M^D =
\Psi_1^{(W)} + \Psi_1^{(E)},\nonumber \\
 &&{}\nonumber \\
 &&{} \nonumber \\
 \Psi_1^{(W)} &=& - {1\over {\sqrt{2}}}\, \Big({1\over {(1 + n)^2}}\,
 {\hat {\bar n}}_{(a)}\, {\hat {\bar \epsilon}}_{(+)(b)}\, {\bar W}_{\tau
 r\tau s}\, {}^3{\bar e}^r_{(a)}\, {}^3{\bar e}^s_{(b)} +\nonumber \\
 &+& {1\over {1 + n}}\, \Big[1 - {{\sqrt{\sum_e\, {\bar n}^2_{(e)}}}\over {1 + n}}\Big]\,
 {\hat {\bar n}}_{(a)}\, {\hat {\bar \epsilon}}_{(+)(c)}
 {\hat {\bar n}}_{(b)}\, {\bar W}_{\tau ruv}\, {}^3{\bar e}^r_{(a)}\, {}^3{\bar
 e}^u_{(b)}\, {}^3{\bar e}^v_{(c)} \Big) -\nonumber \\
  &-& \sgn\, {{4\pi G}\over {c^3}}\, \Big[1 - {{\sqrt{\sum_e\, {\bar n}^2_{(e)}}}\over {1 + n}}\,
 (1 - {{\sqrt{\sum_e\, {\bar n}^2_{(e)}}}\over {1 + n}})\Big]\,
 {\cal T}_{12},\nonumber \\
  &&{}\nonumber \\
 &&{}\nonumber \\
 \Psi_1^{(E)} &=& - {{\sgn}\over 2}\, \Big[1 - {{\sqrt{\sum_e\, {\bar n}^2_{(e)}}}\over {1 + n}}\,
 (1 - {{\sqrt{\sum_e\, {\bar n}^2_{(e)}}}\over {1 + n}})\Big]\,
 {\cal E}_{12} \cir 0,
  \end{eqnarray*}

 \medskip

  \begin{eqnarray*}
 \Psi_2 &=& {1\over 2}\, {}^4C_{ABCD}\, K^A\, L^B\, (K^C\, L^D - M^C\,
 M^{*D}) = \Psi_2^{(W)} + \Psi_2^{(E)},\nonumber \\
 &&{}\nonumber \\
 &&{}\nonumber \\
 \Psi_2^{(W)} &=& {1\over {(1 + n)^2}}\, {\hat {\bar n}}_{(a)}\, {\hat {\bar n}}_{(b)}\,
 {\bar W}_{\tau r\tau s}\, {}^3{\bar e}^r_{(a)}\, {}^3{\bar e}^s_{(b)} +\nonumber \\
 &+& {1\over {1 + n}}\, {\hat {\bar n}}_{(a)}\, {\hat {\bar \epsilon}}_{(+)(b)}\,
 {\hat {\bar \epsilon}}_{(-)(c)}\, {\bar W}_{\tau ruv}\, {}^3{\bar e}^r_{(a)}\,
 {}^3{\bar e}^u_{(b)}\, {}^3{\bar e}^v_{(c)} +\nonumber \\
  &+& {{4\pi G}\over {c^3}}\, {{\sgn}\over
 {(1 + n)^2}}\, \Big[ (1 + n)\, ({\cal T}_{00} + {\cal T}_{22}) +
 \sgn\, T\Big],\nonumber \\
  &&{}\nonumber \\
 &&{}\nonumber \\
 \Psi_2^{(E)} &=& {{\sgn}\over {(1 + n)^2}}\, \Big[{1\over 2}\, (1 + n)\, ({\cal E}_{00} + {\cal E}_{22})
 + {{\sgn}\over 6}\, {\cal E}\Big] \cir 0,
  \end{eqnarray*}

 \medskip

  \begin{eqnarray*}
 \Psi_3 &=& {}^4C_{ABCD}\, K^A\, L^B\, M^{*C}\, L^D = \Psi_3^{(W)} + \Psi_3^{(E)},\nonumber \\
 &&{}\nonumber \\
 &&{}\nonumber \\
 \Psi_3^{(W)} &=& {1\over {\sqrt{2}}}\, \Big({1\over {(1 + n)^2}}\,
 {\hat {\bar n}}_{(a)}\, {\hat {\bar \epsilon}}_{(-)(b)}\, {\bar W}_{\tau
 r\tau s}\, {}^3{\bar e}^r_{(a)}\, {}^3{\bar e}^s_{(b)} -\nonumber \\
 &-& {1\over {1 + n}}\, (1 + {{\sqrt{\sum_e\, {\bar n}^2_{(e)}}}\over {1 + n}})\,
 {\hat {\bar n}}_{(a)}\, {\hat {\bar n}}_{(b)}\, {\hat {\bar
 \epsilon}}_{(-)(c)}\, {\bar W}_{\tau ruv}\, {}^3{\bar e}^r_{(a)}\,
 {}^3{\bar e}^u_{(b)}\, {}^3{\bar e}^v_{(c)} \Big) +\nonumber \\
  &+& \sgn\, {{4\pi G}\over
 {c^3}}\, \Big[{\cal T}^*_{01} + 1 - {{\sum_e\, {\bar n}^2_{(e)}}
 \over {(1 + n)^2}})\, {\cal T}^*_{12}\Big],\nonumber \\
  &&{}\nonumber \\
 &&{}\nonumber \\
 \Psi_3^{(E)} &=& {{\sgn}\over 2}\, \Big[{\cal E}^*_{01} + (1 - {{\sum_e\, {\bar n}^2_{(e)}}
 \over {(1 + n)^2}})\, {\cal E}^*_{12}\Big] \cir 0,
  \end{eqnarray*}

 \medskip

  \bea
 \Psi_4 &=& {}^4C_{ABCD}\, M^{*A}\, L^B\, M^{*C}\, L^D = \Psi_4^{(W)} + \Psi_4^{(E)},\nonumber \\
 &&{}\nonumber \\
 &&{}\nonumber \\
 \Psi_4^{(W)} &=& {1\over {2(1 + n)^2}}\, {\hat {\bar \epsilon}}_{(-)(a)}\,
 {\hat {\bar \epsilon}}_{(-)(b)}\, {\bar W}_{\tau r\tau s}\, {}^3{\bar e}^r_{(a)}\,
 {}^3{\bar e}^s_{(b)} -\nonumber \\
 &-& {1\over {1 + n}}\, (1 + {{\sum_e\, {\bar n}^2_{(e)}}\over {(1 + n)^2}})\,
 {\hat {\bar \epsilon}}_{(-)(a)}\, {\hat {\bar n}}_{(b)}\, {\hat
 {\bar \epsilon}}_{(-)(c)}\, {\bar W}_{\tau ruv}\, {}^3{\bar e}^r_{(a)}\,
 {}^3{\bar e}^u_{(b)}\, {}^3{\bar e}^v_{(c)} +\nonumber \\
 &+& {1\over 2}\, (1 + {{\sum_e\, {\bar n}^2_{(e)}}\over {(1 + n)^2}})^2\,\,
 {\hat {\bar n}}_{(a)}\, {\hat {\bar \epsilon}}_{(-)(b)}\, {\hat {\bar n}}_{(c)}\,
 {\hat {\bar \epsilon}}_{(-)(d)}\, {\bar W}_{rsuv}\, {}^3{\bar e}^r_{(a)}\,
 {}^3{\bar e}^s_{(b)}\, {}^3{\bar e}^u_{(c)}\, {}^3{\bar e}^v_{(d)} +\nonumber \\
  &+& \sgn\, {{2\pi G}\over {c^3}}\, (1 - {{\sum_e\,
 {\bar n}^2_{(e)}}\over {(1 + n)^2}})\, {\cal T}^*_{02},\nonumber \\
  &&{}\nonumber \\
 &&{}\nonumber \\
 \Psi_4^{(E)} &=& {{\sgn}\over 4}\, (1 - {{\sum_e\, {\bar n}^2_{(e)}}\over
 {(1 + n)^2}})\, {\cal E}^*_{02} \cir 0.
  \label{2.6}
 \eea

\bigskip

The Weyl scalars in special bases of null tetrads are used for the Petrov classification
of gravitational fields \cite{3}  (see also Refs.\cite{11,12,13})\footnote{Given the null tetrads I-(4.5) with a given fixed null vector $L^A$, we can define the following family ($A > 0$, B complex, C real are arbitrary) of sets of null tetrads preserving $L^A$: $L^{{'}A} = A\, L^A$, $K^{{'} A} = A^{-1}\, (K^A + B\, B^*\, L^A + B^*\, M^A + B\, M^{*\, A})$, $M^{{'}A} = e^{i\, C}\, (M^A + B\, L^A)$.}.

\subsection{The Weyl Invariants}

The four eigenvalues of the 4-Weyl tensor are the following {\it four 4-scalars invariant functionals} (they do not depend on the choice of the null tetrads and one has $I^3 \not= 6\, J^2$ when the 4-Weyl tensor is not in an algebraically special case of the Petrov classification) \cite{3,11,12,13}

\bea
w_1 &=& \mathrm{Tr} \, ({}^4g\, {}^4C\, {}^4g\, {}^4C),\qquad
w_2 = \mathrm{Tr} \, ({}^4g\, {}^4C\, \epsilon\, {}^4C), \nonumber \\
w_3 &=& \mathrm{Tr} \, ({}^4g\, {}^4C\, {}^4g\, {}^4C\, {}^4g\,
{}^4C), \qquad
w_4 = \mathrm{Tr} \, ({}^4g\, {}^4C\, {}^4g\, {}^4C\, \epsilon\,
{}^4C),\nonumber \\
 &&{}\nonumber \\
 w_1 + i\, w_2 &=& 2\, {}^4{\tilde C}_{ABCD}\, {}^4{\tilde C}^{ABCD} =
 I = 2\, \Psi_0\, \Psi_4 - 8\, \Psi_1\, \Psi_3 + 6\,
\Psi_2^2,\nonumber \\
 w_3 + i\, w_4 &=& 2\, {}^4{\tilde C}_{ABCD}\, {}^4{\tilde C}^{CD}{}_{EF}\,
 {}^4{\tilde C}^{EFAB} = J  = 6\, det\,  \left( \begin{array}{lll} \Psi_0 & \Psi_1 &
\Psi_2 \\ \Psi_1 & \Psi_2& \Psi_3 \\ \Psi_2 & \Psi_3 & \Psi_4
\end{array} \right) = 6\, det\,  \left( \begin{array}{lll} \Psi_0 & \Psi_1 &
\Psi_2 \\ \Psi_1 & \Psi_2& \Psi_3 \\ \Psi_2 & \Psi_3 & \Psi_4
\end{array} \right) ,\nonumber \\
 &&{}
 \label{2.7}
\eea

\noindent where the tensor ${}^4{\tilde C}_{ABCD}$ has the following expression (see Ref.\cite{3}; $\epsilon_{ABCD}$ is the Levi-Civita tensor)

\begin{eqnarray*}
 2\,  {}^4{\tilde C}_{ABCD} &=&  \Big({}^4C_{ABCD} + {i\over 2}\, \eta_{ABEF}\,
 {}^4C^{EF}{}_{CD}\Big) = {1\over 2}\, \Big({}^4C_{ABCD} + {i\over 2}\, {{\sgn\,
 \epsilon_{ABEF}}\over {(1 + n)\, \sqrt{{}^3g}}}\, {}^4C^{EF}{}_{CD}\Big) =\nonumber \\
 &=& \Psi_0\, U_{AB}\, U_{CD} + \Psi_1\, (U_{AB}\, W_{CD} + U_{CD}\, W_{AB}) +\nonumber \\
 &+& \Psi_2\, (U_{AB}\, V_{CD} + U_{CD}\, V_{AB} + W_{AB}\, W_{CD}) +\nonumber \\
 &+& \Psi_3\, (V_{AB}\, W_{CD} + V_{CD}\, W_{AB}) + \Psi_4\, V_{AB}\, V_{CD},
 \end{eqnarray*}

 \bea
 &&U_{AB} = M_A^*\, L_B - M_B^*\, L_A,\qquad V_{AB} = K_A\, M_B - K_B\, M_A,\nonumber \\
 &&W_{AB} = M_A\, M_B^* - M_B\, M_A^* + L_A\, K_B - L_B\, K_A.
 \label{2.8}
 \eea

\medskip

Therefore, like for the Weyl scalars we get the following expression for the Weyl eigenvalues ($k = 1,2,3,4$)

\beq
 w_k = {\bar w}_k + w_k^{(E)} \cir {\bar w}_k,
 \label{2.9}
 \eeq

\noindent with ${\bar w}_h$ Hamiltonian 4-scalar functions of the tidal and inertial gauge variables of the York basis.

\section{The Electric and Magnetic Components of the Weyl Tensor with Respect to the
Congruence of the Eulerian Observers}

The 10 components of the 4-Weyl tensor may be described by the 5
magnetic and the 5 electric components of the 4-Weyl tensor with
respect to a time-like vector field $V = V^{\mu}\,
\partial_{\mu} = V^A\, \partial_A$ (${}^4g^{AB}\, E^{(V)}_{AB} =
{}^4g^{AB}\, H^{(V)}_{AB} = 0$ are implied by ${}^4g^{AC}\,
{}^4C_{ABCD} = 0$)

\medskip

\bea
 E^{(V)}_{AB} &=& E^{(V)}_{BA} = {}^4C_{AEBF}\, V^E\, V^F\qquad
E^{(V)}_{AB}\, V^B = 0,\quad {}^4g^{AB}\, E^{(V)}_{AB} = 0,
\nonumber \\
 &&{}\nonumber \\
 H^{(V)}_{AB} &=& H^{(V)}_{BA} = {}^*{}^4C_{AEBF}\, V^E\, V^F\qquad
H^{(V)}_{AB}\, V^B = 0,\quad {}^4g^{AB}\, H^{(V)}_{AB} = 0,\nonumber \\
 &&{}\nonumber \\
 &&{}\nonumber \\
 {}^4C^{AB}{}_{CD} &=& {}^4g^{AU}\, {}^4g^{BV}\, {}^4C_{UVCD} =\nonumber \\
 &=& \Big[(\delta^A_E\, \delta^B_F - \delta^B_E\, \delta^A_F)\,
 (\delta^G_C\, \delta^H_D - \delta^G_D\, \delta^H_C) -
 \eta^{AB}{}_{EF}\, \eta^{GH}{}_{CD}\Big]\, V^E\, V_G\, E^{(V)
 F}{}_h -\nonumber \\
 &-& \Big[\eta^{AB}{}_{EF}\, (\delta^G_C\, \delta^H_D - \delta^G_D\,
 \delta^H_C) + (\delta^A_E\, \delta^B_F - \delta^B_E\, \delta^A_F)\,
 \eta^{GH}{}_{CD}\Big]\, V^E\, V_G\, H^{(V) F}{}_H,\nonumber \\
 &&{}
 \label{3.1}
 \eea

\noindent where the {\it dual} has the definition ${}^*{}^4C_{ABCD}
= {1\over 2}\, \eta_{AB}{}^{EF}\, {}^4C_{EFCD}$ (see Appendix A).

\bigskip

With every 3+1 splitting of space-time there is the associated
 congruence of the Eulerian observers with the unit normal
$l^{\mu}(\tau, \vec \sigma) = \Big(z^{\mu}_A\, l^A\Big)(\tau, \vec
\sigma)$ to the 3-spaces as unit 4-velocity. The world-lines of
these observers are the integral curves of the unit normal and in
general are not geodesics. In adapted radar 4-coordinates the
contro-variant ($l^A(\tau, \vec \sigma)$, ${}^4{\buildrel \circ
\over {\bar E}}^A_{(a)}(\tau, \vec \sigma)$) and covariant
($l_A(\tau, \vec \sigma)$, ${}^4{\buildrel \circ \over {\bar
E}}_{(a)A}(\tau, \vec \sigma)$) orthonormal tetrads carried by the
Eulerian observers are given in Eqs.I-(2.4).

\bigskip

For the congruence of Eulerian observers  we have $V^A = l^A$ and on-shell we
get\medskip

\begin{eqnarray*}
 E^{(l)}_{\tau\tau} &\cir&
 {\bar{n}_{(a)} \, \bar{n}_{(b)} \over (1+n)^2} \,  {}^3\bar{e}_{(a)}^r \, {}^3\bar{e}_{(b)}^s \,\bar{C}_{\tau r \tau s},
 \end{eqnarray*}

\begin{eqnarray*}
 E^{(l)}_{\tau r} &=& E^{(l)}_{r\tau} \cir
  {\bar{n}_{(a)} \, \bar{n}_{(b)} \over (1+n)^2} \,  {}^3\bar{e}_{(a)}^u \, {}^3\bar{e}_{(b)}^v  \, \bar{C}_{\tau u r v} +  {\bar{n}_{(a)} \over (1+n)^2} \, {}^3\bar{e}_{(a)}^u \, \bar{C}_{\tau r \tau u},\nonumber \\
\end{eqnarray*}

\begin{eqnarray*}
 E^{(l)}_{rs} &\cir&
 {1\over (1+n)^2} \, \bar{C}_{\tau r \tau s} + {\bar{n}_{(a)} \over (1+n)^2} \, {}^3\bar{e}_{(a)}^u \, \left( \bar{C}_{\tau rsu} + \bar{C}_{\tau sru} \right) + {\bar{n}_{(a)} \, \bar{n}_{(b)} \over (1+n)^2} \, {}^3\bar{e}_{(a)}^{u} \, {}^3\bar{e} _{(b)}^v \, \bar{C}_{rusv},
 \end{eqnarray*}

 \begin{eqnarray*}
 H^{(l)}_{\tau\tau} &\cir&
 {\bar{n}_{(a)} \, \bar{n}_{(b)} \over 2 \, (1+n)^2} \, \eta_{\tau ruv} \, {}^3{\bar e}^u_{(c)}\, {}^3{\bar e}^n_{(c)}
 \, {}^3{\bar e}^v_{(d)}\, {}^3{\bar e}^m_{(d)}\, {}^3\bar{e}_{(a)}^r \, {}^3\bar{e}_{(b)}^s \, \bar{C}_{\tau smn},\nonumber \\
\end{eqnarray*}

\begin{eqnarray*}
 H^{(l)}_{\tau r} &=& H^{(l)}_{r\tau} \cir
 {1\over4} \, \Big[ \eta_{\tau e u v} \, {{\bar n}_{(a)} \over (1+n)^2} \, {}^3\bar{e}_{(a)}^e \,
{}^3{\bar e}^u_{(c)}\, {}^3{\bar e}^m_{(c)} \, {}^3{\bar e}^v_{(d)}\, {}^3{\bar e}^n_{(d)}\,
 \left( \bar{C}_{\tau rmn} + \bar{C}_{r w m n} \, \bar{n}_{(b)} \, {}^3\bar{e}^w_{(b)} \right) -\nonumber \\
&-& {\bar{n}_{(a)} \over (1+n)^2} \, {}^3\bar{e}_{(a)}^e \, \eta_{\tau rmn} \,
{}^3{\bar e}^m_{(c)}\, {}^3{\bar e}^u_{(c)}\, \left( 2 \, {\bar{n}_{(b)} \over (1+n)^2} \, \bar{C}_{\tau u \tau e} +
 {}^3{\bar e}^n_{(d)}\, {}^3{\bar e}^v_{(d)}\, \bar{C}_{\tau euv} \right) +-\nonumber \\
&-& 2 \, {\bar{n}_{(a)} \, \bar{n}_{(b)} \over (1+n)^2} \, {}^3\bar{e}_{(a)}^s \, {}^3\bar{e}_{(b)}^e \, \eta_{\tau mrs} \, {}^3{\bar e}^m_{(c)}\, {}^3{\bar e}^u_{(c)}\, \left( \bar{C}_{\tau u \tau e} + \bar{C}_{\tau euv} \, \bar{n}_{(c)} \, {}^3\bar{e}^v_{(c)} \right) \Big],\nonumber \\
\end{eqnarray*}

\bea
 H^{(l)}_{rs} &\cir& {1\over 4} {1\over (1+n)^2} \,
 {}^3{\bar e}^v_{(c)}\, {}^3{\bar e}^m_{(c)}\, {}^3{\bar e}^u_{(d)}\, {}^3{\bar e}^n_{(d)}\,
  \Big[ \eta_{\tau ruv} \left( \bar{C}_{\tau smn} + \bar{C}_{mnsw} \bar{n}_{(a)}\,{}^3\bar{e}_{(a)}^w \right) +
  \nonumber \\
&+& \eta_{\tau suv} \, \left( \bar{C}_{\tau rmn} + \bar{C}_{mnrw} \, \bar{n}_{(a)} \, {}^3\bar{e}_{(a)}^w \right) \Big].
 \label{3.2}
 \eea

\medskip

One could also find the electric and magnetic components of the 4-Weyl tensor with respect to the
 skew congruence with unit 4-velocity $v^{\mu}(\tau, \vec
\sigma) = \Big({{z^{\mu}_{\tau}}\over {\sqrt{(1 + n)^2 - \sum_c\, {\bar n}^2_{(c)}}
 }}\, v^A\Big)(\tau, \vec \sigma)$ (in general
it is not surface-forming, i.e. it has a non-vanishing vorticity)
associated with each 3+1 splitting of the space-time.
The observers of the skew congruence have the world-lines (integral
curves of the 4-velocity) defined by $\sigma^r = const.$ for every
$\tau$. When there is a perfect fluid with unit time-like 4-velocity
$U^A(\tau, \vec \sigma)$, there is also the congruence of the
time-like flux curves: in general it is not surface-forming and it
is independent from the previous two congruences. See Ref.\cite{14} for
the description of these two congruences.

\section{The Dirac Observables of Canonical Gravity versus the Weyl Eigenvalues as Possible 4-scalar Bergmann
Observables}

In Section II of I we reviewed the formulation of canonical ADM tetrad gravity in the York basis I-(28) developed in Refs. \cite{4,5,6,7,15,16} (see Ref. \cite{17} for canonical ADM metric gravity).
\medskip

In this formulation the only constraints not contained in the York basis are the super-Hamiltonian (${\cal H}(\tau, \vec \sigma) \approx 0$) and the super-momentum (${\cal H}_{(a)}(\tau, \vec \sigma) \approx 0$) ones. As it is clear from Eqs. (3.41) and (3.44) of Ref.\cite{5}, these constraints are elliptic partial differential equations inside the 3-spaces $\Sigma_{\tau}$ for the unknowns $\tilde \phi (\tau, \vec \sigma)$ and $\pi_i^{(\theta)}(\tau, \vec \sigma)$ (or for the shear components $\sigma_{(a)(b)}{|}_{a \not= b}$ defined after Eq.I-(2.11)) depending on the canonical variables $\pi_{\tilde \phi}(\tau, \vec \sigma)$, $\theta^i(\tau, \vec \sigma)$ (the primary inertial gauge variables), and $R_{\bar a}(\tau, \vec \sigma)$, $\Pi_{\bar a}(\tau, \vec \sigma)$ (the tidal variables), but not on the lapse and shift functions $1 + n(\tau, \vec \sigma)$, ${\bar n}_{(a)}(\tau, \vec \sigma)$ (the secondary inertial gauge variables, to be determined by the time-constancy of the gauge fixings for the primary ones).

\bigskip

If in a suitable function space one would be able to find a solution

\bea
 \tilde \phi(\tau, \vec \sigma) &\approx& h(\theta^j, \pi_{\tilde \phi}, R_{\bar a}, \Pi_{\bar a} | \tau, \vec \sigma),\nonumber \\
 \pi_i^{(\theta)}(\tau, \vec \sigma) &\approx& h_i(\theta^j, \pi_{\tilde \phi}, R_{\bar a}, \Pi_{\bar a} | \tau, \vec \sigma),
 \label{4.1}
 \eea

\noindent of these elliptic PDE equations, then one could find a Shanmugadhasan canonical transformation to a
final canonical basis adapted to all the 14 first class constraints of ADM tetrad gravity \footnote{The constraints
$\hat \phi(\tau, \vec \sigma) \approx 0$ and ${\hat \pi}_i^{(\theta)}(\tau, \vec \sigma) \approx 0$ are in strong involution, namely they have exactly zero Poisson brackets (differently from the original constraints which are only in weak involution), because they are explicitly solved in certain canonical variables of the York basis.} (from now on we consider only the case without matter to simplify the discussion)

\begin{eqnarray*}
 &&\begin{minipage}[t]{4 cm}
\begin{tabular}{|ll|ll|l|l|l|} \hline
$\varphi_{(a)}$ & $\alpha_{(a)}$ & $n$ & ${\bar n}_{(a)}$ &
$\theta^r$ & $\tilde \phi$ & $R_{\bar a}$\\ \hline
$\pi_{\varphi_{(a)}} \approx0$ &
 $\pi^{(\alpha)}_{(a)} \approx 0$ & $\pi_n \approx 0$ & $\pi_{{\bar n}_{(a)}} \approx 0$
& $\pi^{(\theta )}_r$ & $\pi_{\tilde \phi}$ & $\Pi_{\bar a}$ \\
\hline
\end{tabular}
\end{minipage}\nonumber \\
  &&{}\nonumber \\
 &&{\longrightarrow \hspace{.2cm}} \
\begin{minipage}[t]{4 cm}
\begin{tabular}{|ll|ll|l|l|l|} \hline
$\varphi_{(a)}$ & $\alpha_{(a)}$ & $n$ & ${\bar n}_{(a)}$ &
${\hat \theta}^r$ & $\hat \phi \approx 0$ & ${\hat R}_{\bar a}$\\ \hline
$\pi_{\varphi_{(a)}} \approx0$ &
 $\pi^{(\alpha)}_{(a)} \approx 0$ & $\pi_n \approx 0$ & $\pi_{{\bar n}_{(a)}} \approx 0$
& ${\hat \pi}^{(\theta )}_r \approx 0$ & ${\hat \pi}_{\hat \phi}$ & ${\hat\Pi}_{\bar a}$ \\
\hline
\end{tabular}
\end{minipage}\nonumber \\
 \end{eqnarray*}

\bea
 \hat \phi(\tau, \vec \sigma) &=&  \tilde \phi(\tau, \vec \sigma) - h(\theta^j, \pi_{\tilde \phi}, R_{\bar a}, \Pi_{\bar a} | \tau, \vec \sigma) \approx 0,\qquad
 {\hat \pi}_{\hat \phi}(\tau, \vec \sigma) =
  f(\theta^j, \pi_{\tilde \phi}, R_{\bar a}, \Pi_{\bar a} | \tau, \vec \sigma),\nonumber \\
 {\hat \theta}^i(\tau, \vec \sigma) &=&
    f^i(\theta^j, \pi_{\tilde \phi}, R_{\bar a}, \Pi_{\bar a} | \tau, \vec \sigma),\qquad
 {\hat \pi}_i^{(\theta)}(\tau, \vec \sigma) = \pi_i^{(\theta)}(\tau, \vec \sigma) -
 h_i(\theta^j, \pi_{\tilde \phi}, R_{\bar a}, \Pi_{\bar a} | \tau, \vec \sigma) \approx 0,
 \nonumber \\
  {\hat R}_{\bar a}(\tau, \vec \sigma) &=&
  k_{\bar a}(\theta^j, \pi_{\tilde \phi}, R_{\bar a}, \Pi_{\bar a} | \tau, \vec \sigma),\qquad
  {\hat \Pi}_{\bar a}(\tau, \vec \sigma) =
    g_{\bar a}(\theta^j, \pi_{\tilde \phi}, R_{\bar a}, \Pi_{\bar a} | \tau, \vec \sigma).\nonumber \\
  {}&&
 \label{4.2}
 \eea

\medskip

In this final canonical basis ${\hat R}_{\bar a}(\tau, \vec \sigma)$ and ${\hat \Pi}_{\bar a}(\tau, \vec \sigma)$
would be two canonical pairs of true DO's  (completely invariant under all the Hamiltonian gauge transformations)
describing the real tidal physical degrees of freedom of the gravitational field. Like the variables of the York basis these DO's would be  3-scalars inside the 3-spaces $\Sigma_{\tau}$ and  4-scalars of the space-time (invariant under
passive diffeomorphisms, i.e. under the group of gauge transformations of the Lagrangian theory giving rise to the notion of general covariance).

\medskip
${\hat \theta}^i(\tau, \vec \sigma)$ and ${\hat \pi}_{\hat \phi}(\tau, \vec \sigma)$ would be the final primary inertial gauge variables describing the freedom in the choice of the tangent vectors to the spatial coordinate lines and in the choice of a variable like the York time describing the gauge part of non-Euclidean nature of the 3-spaces.
The secondary inertial gauge variables $1 + n(\tau, \vec \sigma)$ and ${\bar n}_{(a)}(\tau, \vec \sigma)$ would remain unchanged, because Eqs.(\ref{4.1}) do not depend on them.

Since the solved constraints $\hat \phi = \tilde \phi - h(...) \approx o$ and ${\hat \pi}_i^{(\theta)} = \pi_i^{(\theta)} - h_i(...) \approx 0$ are linear in the variables $\tilde \phi$ and $\pi_i^{(\theta)}$, it is highly probable that the primary gauge variables remain unchanged: ${\hat \pi}_{\hat \phi} = f(...) = \pi_{\tilde \phi}$ and ${\hat \theta}^i = f^i(...) = \theta^i$. If this is true, only the tidal variables (and the matter when present) would be modified by the final Shanmugadhasan canonical transformation.
\bigskip

The Dirac Hamiltonian I-(2.14) would be replaced by a new Hamiltonian

\bea
  H^{(F)}_D&=& {1\over c}\, {\hat E}^{(F)}_{ADM} + \int d^3\sigma\, \Big[ {\cal F}\,
\hat \phi - {\cal F}^i\, {\hat \pi}_i^{(\theta)}\Big](\tau ,
\vec \sigma)+ \lambda_r(\tau )\, {\hat P}^r_{ADM} +\nonumber \\
 &+&\int d^3\sigma\, \Big[\lambda_n\, \pi_n + \lambda_{
{\bar n}_{(a)}}\, \pi_{{\bar n}_{(a)}} + \lambda_{\varphi_{(a)}}\,
\pi_{ \varphi_{(a)}} + \lambda_{\alpha_(a)}\,
\pi^{(\alpha)}_{(a)}\Big](\tau , \vec \sigma ),
 \label{4.3}
 \eea

\noindent in which the weak ADM energy ${\hat E}_{ADM}^{(F)}$ depends on the final tidal variables and on the final primary inertial gauge variables but not on the secondary ones. The functions ${\cal F}$ and ${\cal F}^i$ in front of the Abelianized form of the secondary first-class constraints (replacing the lapse and shift functions) must be determined by the following comparison of the Hamilton equations before and after the final canonical transformation
($H_D$ is the old Dirac Hamiltonian of Eqs. I-(2.14); see Ref.\cite{5} for the explicit form of the Hamilton equations)

\bea
 &&{\cal F}^i({\hat \theta}^j, {\hat \pi}_{\hat \phi}, {\hat R}_{\bar a}, {\hat \Pi}_{\bar a}, n, {\bar n}_{(a)} | \tau, \vec \sigma) = \{ {\hat \theta}^i(\tau, \vec \sigma), H_D^{(F)} \} \cir \partial_{\tau}\, {\hat \theta}^i(\tau, \vec \sigma) =\nonumber \\
 &&= \partial_{\tau}\, f^i(\theta^j, \pi_{\tilde \phi}, R_{\bar a}, \Pi_{\bar a} | \tau, \vec \sigma) \cir
 \{ f^i(\theta^j, \pi_{\tilde \phi}, R_{\bar a}, \Pi_{\bar a} | \tau, \vec \sigma), H_D \} =\nonumber \\
 &&= {\tilde f}^i(\theta^j, \pi_{\tilde \phi}, R_{\bar a}, \Pi_{\bar a}, n, {\bar n}_{(a)} | \tau, \vec \sigma)
 =\nonumber \\
 &&= {\hat {\cal G}}^i({\hat \theta}^j, {\hat \pi}_{\hat \phi}, {\hat R}_{\bar a}, {\hat \Pi}_{\bar a}, n, {\bar n}_{(a)} | \tau, \vec \sigma),\nonumber \\
 {}&&\nonumber \\
  &&{\cal F}({\hat \theta}^j, {\hat \pi}_{\hat \phi}, {\hat R}_{\bar a}, {\hat \Pi}_{\bar a}, n, {\bar n}_{(a)} | \tau, \vec \sigma) = \{ {\hat \pi}_{\hat \phi}(\tau, \vec \sigma), H_D^{(F)} \} \cir \partial_{\tau}\, {\hat \pi}_{\hat \phi}(\tau, \vec \sigma) =\nonumber \\
 &&= \partial_{\tau}\, f(\theta^j, \pi_{\tilde \phi}, R_{\bar a}, \Pi_{\bar a} | \tau, \vec \sigma) \cir
 \{ f(\theta^j, \pi_{\tilde \phi}, R_{\bar a}, \Pi_{\bar a} | \tau, \vec \sigma), H_D \} =\nonumber \\
 &&= {\tilde f}(\theta^j, \pi_{\tilde \phi}, R_{\bar a}, \Pi_{\bar a}, n, {\bar n}_{(a)} | \tau, \vec \sigma)
 =\nonumber \\
 &&= {\hat {\cal G}}({\hat \theta}^j, {\hat \pi}_{\hat \phi}, {\hat R}_{\bar a}, {\hat \Pi}_{\bar a}, n, {\bar n}_{(a)} | \tau, \vec \sigma).
 \label{4.4}
 \eea

If we give primary gauge fixings ${\hat \theta}^i(\tau, \vec \sigma) - v^i(\tau, \vec \sigma) \approx 0$ and ${\hat \pi}_{\hat \phi}(\tau, \vec \sigma) - v(\tau, \vec \sigma) \approx 0$ with $v^i(\tau, \vec \sigma)$ and $v(\tau, \vec \sigma)$ numerical functions, then, due to Eqs.(\ref{4.3}) and (\ref{4.4}), the secondary gauge fixings for the determination of the lapse and shift functions are ${\cal F}^i(\tau, \vec \sigma) - \partial_{\tau}\, v^i(\tau, \vec \sigma) \approx 0$ and ${\cal F}(\tau, \vec \sigma) - \partial_{\tau}\, v(\tau, \vec \sigma) \approx 0$. These are elliptic PDE inside the 3-space $\Sigma_{\tau}$. Instead the DO's satisfy hyperbolic PDE requiring Cauchy data on an initial Cauchy surface $\Sigma_{\tau_o}$.

Therefore the final DO's would also be  BO's, since they would be 4-scalars uniquely predictable from the Cauchy data.

\bigskip

Regarding the Weyl eigenvalues, Eqs. (\ref{2.7}) and (\ref{2.9}) show that their 4-scalar Hamiltonian expressions in the York basis depends on both the primary and secondary gauge variables. Therefore, most probably their Hamiltonian expression in the final Shanmugadhasan canonical basis will depend on the lapse and shift functions. If this will turn out to be correct, then the Weyl eigenvalues would not be BO's, contrary to the old proposal of Bergmann and Komar \cite{8}.

\bigskip

Instead in Refs. \cite{6,7} it was shown that the Hamiltonian expression (\ref{2.9}) of the four Weyl eigenvalues ($w_k \cir {\bar w}_k$) can be used to define a special family of gauges, whose primary gauge fixings have the form of coordinate conditions for the radar 4-coordinates $\sigma^A = (\tau, \sigma^r)$

\beq
 \chi^A(\tau, \vec \sigma) = \sigma^A - {\cal U}^A({\bar w}_k(\tau, \vec \sigma)) \approx 0.
 \label{4.5}
 \eeq

These equations determine the primary gauge variables and their $\tau$-constancy, $\partial_{\tau}\, \chi^A(\tau, \vec \sigma) \approx 0$, determines the lapse and shift functions. Eqs.(\ref{4.5}) imply that the radar 4-coordinates $\sigma^A$ (replacing the usual world 4-coordinates $x^{\mu}$ in our approach) labeling the {\it mathematical} points of the space-time 4-manifold may acquire a {\it physical} meaning in terms of the gravitational field. The abstract mathematical points can be identified as physically individuated point-events. Therefore, the gravitational field, which describes the metric structure of the space-time by definition, can also be used a posteriori to give a physical meaning to the 4-manifold carrying that metric structure due to Einstein's equations. This is not possible for Minkowski and Galileo space-times, which remain as absolutely given mathematical 4-manifolds.

\bigskip

Let us come back to the DO's.
In a completely fixed gauge $G$ of the York basis the tidal variables $R_{\bar a}$, $\Pi_{\bar a}$, would become the non-canonical DO's of that gauge. Then the evaluation of Darboux canonical basis for the Dirac brackets would identify the real canonical DO's $R^{(G)}_{\bar a}$, $\Pi^{(G)}_{\bar a}$, of that gauge: they should be connected by a canonical transformation to the DO's ${\hat R}_{\bar a}$, ${\hat \Pi}_{\bar a}$, of the final Shanmugadhasan canonical basis (\ref{4.2}).

\bigskip

As a step towards the implementation of this comparison in Appendix B we review what is known about the 3-orthogonal Schwinger time gauges and about arbitrary gauges near the 3-orthogonal ones for the York basis and their HPM linearization.\medskip

In the next Section these linearized results will be used to give the linearized expression of the Weyl scalars and of the Weyl eigenvalues.

\section{The Linearized Weyl Scalars in Arbitrary Gauges
near the 3-Orthogonal Gauges in Absence of Matter}

Let us consider the HPM linearization of the 4-Weyl tensor and of the Weyl scalars in absence of matter with the formalism of Refs.\cite{14,15} reviewed in Appendix B, where it is shown that the same results can be obtained in the family of 3-orthogonal Schwinger time gauges defined in Eq.(\ref{b1}) and in arbitrary gauges near the 3-orthogonal ones (see Subsections 1 and 2 of Appendix B, respectively). In both cases the inertial gauge variable York time (either gauge fixed or arbitrary) is a first order quantity ${}^3K_{(1)} = {{12 \pi G}\over {c^3}}\, \pi_{(1) \tilde \phi}$.

\bigskip

By using Eqs. I-(3.5), I-(3.6), I-(3.7), (\ref{2.2}) and the solution (\ref{b5}) of the super-momentum constraints  we get \footnote{ As said in paper I the evaluation of the quantities appearing in these equations requires the use of Eqs. I-(2.11), I-(A1), I-(2.12) and I-(2.13). In place of the shear appearing in these equations, here we use $\sigma_{(1)(a)(a)} = - {{8 \pi G}\over {c^3}}\, \sum_{\bar a}\,
\gamma_{\bar aa}\, \Pi_{\bar a}$ and the solution $\sigma_{(1)(a)(b)}{|}_{a \not= b}$ of the
super-momentum constraints given in Eq.(\ref{b5}). Due to Eqs. I-(2.11) and (\ref{b5}) the linearized extrinsic curvature is ${}^3K_{rs} \approx \delta_{rs}\, \Big({1\over 3}\, {}3K_{(1)} - {{8 \pi G}\over {c^3}}\,
\sum_{\bar a}\, \gamma_{\bar ar}\, \Pi_{\bar a}\Big) + {1\over 2}\, (1 - \delta_{rs})\, (\partial_r\, {\bar n}_{(1)(s)} + \partial_s\, {\bar n}_{(1)(r)})$ with trace ${}^3K_{(1)} = {{12 \pi G}\over {c^3}}\, \pi_{(1) \tilde \phi}$.} in absence of matter

\begin{eqnarray*}
 {\bar W}_{(1)rsuv} &=& {\bar {\cal W}}_{(1)rsuv} ={}4R_{(1)rsuv} = {\bar {\cal C}}_{(1)rsuv}\,\, (\cir {}^4C_{(1)rsuv}) = - \sgn\, {}^3R_{(1)rsuv} =\nonumber \\
 &=&  \sgn\, \Big[(\delta_{rv}\, \partial_u - \delta_{ru}\, \partial_v)\, \partial_s\,
 (\Gamma_r^{(1)} + 2\, \phi_{(1)}) +
 (\delta_{su}\, \partial_v - \delta_{sv}\, \partial_u)\, \partial_r\,
 (\Gamma_s^{(1)} + 2\, \phi_{(1)})  \Big],\nonumber \\
 \end{eqnarray*}

\begin{eqnarray*}
 {\bar W}_{(1)\tau ruv} &=& {\bar {\cal W}}_{(1)\tau ruv} ={}4R_{(1)\tau ruv} = {\bar {\cal C}}_{(1)\tau ruv}\,\, (\cir {}^4C_{(1)\tau ruv}) =\nonumber \\
   &=&  \sgn\, (\partial_v\, {}^3K_{(1)ru} -
 \partial_u\, {}^3K_{(1)rv}) =\nonumber \\
 &=& \sgn\, \Big[{1\over 2}\, (1 - \delta_{ru})\, \partial_v\,
 (\partial_r\, {\bar n}_{(1)(u)} + \partial_u\, {\bar n}_{(1)r}))
  - {1\over 2}\, (1 - \delta_{rv})\, \partial_u\,
 (\partial_r\, {\bar n}_{(1)(v)} + \partial_v\, {\bar n}_{(1)r})) +\nonumber \\
 &+& (\delta_{ru}\, \partial_v - \delta_{rv}\, \partial_u)\,
 ({1\over 3}\, {}^3K_{(1)} - {{8 \pi G}\over {c^3}}\, \sum_{\bar a}\, \gamma_{\bar ar}\, \Pi_{\bar a}
  + \partial_r\, {\bar n}_{(1)(r)} -
 {1\over 3}\, \sum_c\, \partial_c\, {\bar n}_{(1)(c)}) \Big],\nonumber \\
 \end{eqnarray*}

\bea
 {\bar W}_{(1)\tau r\tau s} &=& {\bar {\cal W}}_{(1)rsuv} =
 {\bar {\cal C}}_{(1)\tau r\tau s}\,\, (\cir {}4R_{(1)\tau r\tau s} \cir {}^4C_{(1)\tau r\tau s}) =\nonumber \\
   &=&  \sgn\, {}^3R_{(1)rs} =
  \sgn\, \Big[\delta_{rs}\, \triangle\, (\Gamma_r^{(1)} + 2\,
 \phi_{(1)}) + \partial_r\, \partial_s\, (2\, \phi_{(1)} -
 \Gamma_r^{(1)} - \Gamma_s^{(1)}) \Big].
 \label{5.1}
 \eea

\bigskip

By using the solution $\phi_{(1)} \approx - {1\over 4}\, \sum_c\,
{{\partial_c^2}\over {\triangle}}\, \Gamma_c^{(1)}$
of the super-Hamiltonian constraint given in Eq.(\ref{b5}),
 we get that ${\bar W}_{(1)\tau r\tau s}$ and ${\bar W}_{(1)rsuv}$ depend only on $R_{\bar
a}$. Instead ${\bar W}_{(1)\tau ruv}$ depends on $R_{\bar a}$, $\Pi_{\bar a}$ and ${}^3K_{(1)}$, because
we have ${\bar n}_{(1)(a)} \approx {{\partial_a}\over {\triangle}}\, {}^3K_{(1)} + {1\over 2}\,
 {{\partial_a}\over {\triangle}}\, \partial_{\tau}\, \Big(4\,
 \Gamma_a^{(1)} - \sum_c\, {{\partial_c^2}\over {\triangle}}\,
 \Gamma_c^{(1)}\Big)$ due to Eqs.(\ref{b6}). \medskip

Therefore ${\bar W}_{\tau r\tau s}(\tau, \sigma^u)$ and ${\bar
W}_{rsuv}(\tau, \sigma^u)$ have vanishing Poisson bracket among
themselves; the same is true for ${\bar W}_{\tau ruv}(\tau,
\sigma^u)$ with itself. Instead the Poisson bracket of ${\bar
W}_{\tau ruv}(\tau, \sigma^u)$ with either ${\bar W}_{rsuv}(\tau,
\sigma^u)$ or ${\bar W}_{\tau r\tau s}(\tau, \sigma^u)$ have complicated expressions.

\subsection{The Weyl Scalars}

Since Eqs. I-(4.1), I-(4.2) and I-(4.3) imply that the quantities ${\hat {\bar n}}_{(a)} = {\bar n}_{(1)(a)} / \sqrt{\sum_c\, {\bar n}^2_{(c)}}$ and ${\hat {\bar \epsilon}}_{(\pm)(a)} = {1\over {\sqrt{2}}}\,
\Big({\hat {\bar \epsilon}}_{(1)(a)} \pm i\, {\hat {\bar \epsilon}}_{(2)(a)}\Big)$
\footnote{We have ${\hat {\bar \epsilon}}_{(1)(1)} = {\bar n}_{(1)(1)}\, {\bar n}_{(1)(3)} / A$,
${\hat {\bar \epsilon}}_{(1)(2)} = {\bar n}_{(1)(2)}\, {\bar n}_{(1)(3)} / A$,
${\hat {\bar \epsilon}}_{(1)(3)} = - ({\bar n}^2_{(1)(1)} + {\bar n}^2_{(1)(2)}) / A$,
${\hat {\bar \epsilon}}_{(2)(1)} = - {\bar n}_{(1)(2)} / B$,
${\hat {\bar \epsilon}}_{(2)(2)} = {\bar n}_{(1)(1)} / B$, ${\hat {\bar \epsilon}}_{(2)(3)} = 0$, with
$A = \sqrt{({\bar n}^2_{(1)(1)} + {\bar n}^2_{(1)(2)})\, \sum_c\, {\bar n}^2_{(1)(c)}}$ and $B = \sqrt{{\bar n}^2_{(1)(1)} + {\bar n}^2_{(1)(2)}}$.} are of order $O(1)$, for the Weyl scalars $\Psi_A^{(W)}$, $A=0,1,2,3,4$, of Eq.(\ref{2.6}) we get the following linearization $\Psi_{(1)A}^{(W)} = \Psi_{(1)A(R)}^{(W)} + i\, \Psi_{(1)A(I)}^{(W)}
\cir \Psi_{(1)A}$ near the 3-orthogonal gauges

\begin{eqnarray*}
 \Psi_{(1)0}^{(W)} &=& - {1\over 2}\, \Big[{\hat {\bar
 \epsilon}}_{(+)(r)}\, {\hat {\bar
 \epsilon}}_{(+)(s)}\, {\bar W}_{(1)\tau r\tau s} + 2\,
 {\hat {\bar \epsilon}}_{(+)(r)}\, {\hat {\bar n}}_{(u)}\,
 {\hat {\bar \epsilon}}_{(+)(v)}\, {\bar W}_{(1)\tau ruv}
 +\nonumber \\
 &+& {\hat {\bar n}}_{(r)}\, {\hat {\bar
 \epsilon}}_{(+)(s)}\, {\hat {\bar n}}_{(u)}\, {\hat {\bar
 \epsilon}}_{(+)(v)}\, {\bar W}_{(1)rsuv} \Big] =\nonumber \\
 &=& - {{\sgn}\over 4}\, \Big({\hat {\bar \epsilon}}_{(1)(r)}\, {\hat {\bar \epsilon}}_{(1)(s)}
 - {\hat {\bar \epsilon}}_{(2)(r)}\, {\hat {\bar \epsilon}}_{(2)(s)} +
 i\, ({\hat {\bar \epsilon}}_{(1)(r)}\, {\hat {\bar \epsilon}}_{(2)(s)}
 + {\hat {\bar \epsilon}}_{(2)(r)}\, {\hat {\bar \epsilon}}_{(1)(s)})\Big)\nonumber \\
 &&\Big[\delta_{rs}\, \triangle\, (\Gamma_r^{(1)} + 2\,
 \phi_{(1)}) + \partial_r\, \partial_s\, (2\, \phi_{(1)} -
 \Gamma_r^{(1)} - \Gamma_s^{(1)}) +\nonumber \\
 &+& {\hat {\bar n}}_{(u)}\, \Big(\sum_w\, ((1 - \delta_{ru})\, \delta_{uw}\, \partial_s - (1 - \delta_{rs})\, \delta_{sw}\, \partial_u)\,(\partial_r\, {\bar n}_{(1)(w)} + \partial_w\, {\bar n}_{(1)(r)}) +\nonumber \\
 &+& 2\, (\delta_{ru}\, \partial_s - \delta_{rs}\, \partial_u)\,
 ({1\over 3}\, {}^3K_{(1)} - {{8 \pi G}\over {c^3}}\, \sum_{\bar a}\, \gamma_{\bar ar}\, \Pi_{\bar a}
  + \partial_r\, {\bar n}_{(1)(r)} -
 {1\over 3}\, \sum_c\, \partial_c\, {\bar n}_{(1)(c)}) \Big) +\nonumber \\
 &+& {\hat {\bar n}}_{(u)}\, {\hat {\bar n}}_{(v)}\, \Big(
 (\delta_{sv}\, \partial_u - \delta_{uv}\, \partial_s)\, \partial_r\,
 (\Gamma_v^{(1)} + 2\, \phi_{(1)}) +
 (\delta_{ru}\, \partial_s - \delta_{rs}\, \partial_u)\, \partial_v\,
 (\Gamma_r^{(1)} + 2\, \phi_{(1)})  \Big)\Big] =\nonumber \\
 &=& \Psi_{(1)0(R)}^{(W)} + i\, \Psi_{(1)0(I)}^{(W)},
 \end{eqnarray*}

\begin{eqnarray*}
 \Psi_{(1)1}^{(W)} &=&- {1\over {\sqrt{2}}}\, {\hat {\bar
 n}}_{(r)}\, \Big[{\hat {\bar \epsilon}}_{(+)(s)}\, {\bar W}_{(1)\tau r\tau
 s} + {\hat {\bar \epsilon}}_{(+)(u)}\, {\hat {\bar n}}_{(v)}\,
 {\bar W}_{(1)\tau ruv} \Big] =\nonumber \\
 &=& - {{\sgn}\over 2}\, {\hat {\bar n}}_{(r)}\,({\hat {\bar \epsilon}}_{(1)(s)} + i\,
 {\hat {\bar \epsilon}}_{(2)(s)})\, \Big( \delta_{rs}\, \triangle\, (\Gamma_r^{(1)} + 2\,
 \phi_{(1)}) + \partial_r\, \partial_s\, (2\, \phi_{(1)} -
 \Gamma_r^{(1)} - \Gamma_s^{(1)}) +\nonumber \\
 &+& {1\over 2}\, {\hat {\bar n}}_{(u)}\, \Big[
  \sum_w\, ((1 - \delta_{ru})\, \delta_{sw}\, \partial_u - (1 - \delta_{ru})\, \delta_{uw}\, \partial_s)\,(\partial_r\, {\bar n}_{(1)(w)} + \partial_w\, {\bar n}_{(1)(r)}) +\nonumber \\
 &+& 2\, (\delta_{rs}\, \partial_u - \delta_{ru}\, \partial_s)\,
 ({1\over 3}\, {}^3K_{(1)} - {{8 \pi G}\over {c^3}}\, \sum_{\bar a}\, \gamma_{\bar ar}\, \Pi_{\bar a}
  + \partial_r\, {\bar n}_{(1)(r)} -
 {1\over 3}\, \sum_c\, \partial_c\, {\bar n}_{(1)(c)}) \Big]\Big) =\nonumber \\
 &=& \Psi_{(1)1(R)}^{(W)} + i\, \Psi_{(1)1(I)}^{(W)},
 \end{eqnarray*}

\begin{eqnarray*}
 \Psi_{(1)2}^{(W)} &=& {\hat {\bar n}}_{(r)}\, \Big[{\hat {\bar
 n}}_{(s)}\, {\bar W}_{(1)\tau r\tau s} + {\hat {\bar
 \epsilon}}_{(+)(u)}\, {\hat {\bar \epsilon}}_{(-)(v)}\,
 {\bar W}_{(1)\tau ruv} \Big] =\nonumber \\
 &=& \sgn\, {\hat {\bar n}}_{(r)}\, \Big({\hat {\bar n}}_{(s)}\,
 [\delta_{rs}\, \triangle\, (\Gamma_r^{(1)} + 2\,
 \phi_{(1)}) + \partial_r\, \partial_s\, (2\, \phi_{(1)} -
 \Gamma_r^{(1)} - \Gamma_s^{(1)}) ] -\nonumber \\
 &-& {1\over 4}\, [ {\hat {\bar \epsilon}}_{(1)(u)}\, {\hat {\bar \epsilon}}_{(1)(v)}
 + {\hat {\bar \epsilon}}_{(2)(u)}\, {\hat {\bar \epsilon}}_{(2)(v)} -
 i\, ({\hat {\bar \epsilon}}_{(1)(u)}\, {\hat {\bar \epsilon}}_{(2)(v)}
 - {\hat {\bar \epsilon}}_{(2)(u)}\, {\hat {\bar \epsilon}}_{(1)(v)}) ]\nonumber \\
 &&\Big[\sum_w\, ((1 - \delta_{ru})\, \delta_{uw}\, \partial_v - (1 - \delta_{rv})\, \delta_{vw}\, \partial_u)\,(\partial_r\, {\bar n}_{(1)(w)} + \partial_w\, {\bar n}_{(1)(r)}) +\nonumber \\
 &+& 2\, (\delta_{ru}\, \partial_v - \delta_{rv}\, \partial_u)\,
 ({1\over 3}\, {}^3K_{(1)} - {{8 \pi G}\over {c^3}}\, \sum_{\bar a}\, \gamma_{\bar ar}\, \Pi_{\bar a}
  + \partial_r\, {\bar n}_{(1)(r)} -
 {1\over 3}\, \sum_c\, \partial_c\, {\bar n}_{(1)(c)}) \Big]\Big) =\nonumber \\
 &=& \Psi_{(1)2(R)}^{(W)} + i\, \Psi_{(1)2(I)}^{(W)},
\end{eqnarray*}

\begin{eqnarray*}
 \Psi_{(1)3}^{(W)} &=& {1\over {\sqrt{2}}}\, {\hat {\bar n}}_{(r)}\, \Big[
 {\hat {\bar \epsilon}}_{(-)(s)}\, {\bar W}_{(1)\tau r\tau s} - {\hat
 {\bar n}}_{(u)}\, {\hat {\bar \epsilon}}_{(-)(v)}\,
 {\bar W}_{(1)\tau ruv} \Big] =\nonumber \\
 &=& {{\sgn}\over 2}\, {\hat {\bar n}}_{(r)}\,({\hat {\bar \epsilon}}_{(1)(s)} - i\,
 {\hat {\bar \epsilon}}_{(2)(s)})\, \Big( \delta_{rs}\, \triangle\, (\Gamma_r^{(1)} + 2\,
 \phi_{(1)}) + \partial_r\, \partial_s\, (2\, \phi_{(1)} -
 \Gamma_r^{(1)} - \Gamma_s^{(1)}) -\nonumber \\
 &-& {1\over 2}\, {\hat {\bar n}}_{(u)}\, \Big[
  \sum_w\, ((1 - \delta_{ru})\, \delta_{sw}\, \partial_u - (1 - \delta_{ru})\, \delta_{uw}\, \partial_s)\,(\partial_r\, {\bar n}_{(1)(w)} + \partial_w\, {\bar n}_{(1)(r)}) +\nonumber \\
 &+& 2\, (\delta_{rs}\, \partial_u - \delta_{ru}\, \partial_s)\,
 ({1\over 3}\, {}^3K_{(1)} - {{8 \pi G}\over {c^3}}\, \sum_{\bar a}\, \gamma_{\bar ar}\, \Pi_{\bar a}
  + \partial_r\, {\bar n}_{(1)(r)} -
 {1\over 3}\, \sum_c\, \partial_c\, {\bar n}_{(1)(c)}) \Big]\Big) =\nonumber \\
 &=& \Psi_{(1)3(R)}^{(W)} + i\, \Psi_{(1)3(I)}^{(W)},
\end{eqnarray*}

\bea
 \Psi_{(1)4}^{(W)} &=& {1\over 2}\, \Big[{\hat {\bar \epsilon}}_{(-)(r)}\,
 {\hat {\bar \epsilon}}_{(-)(s)}\, {\bar W}_{(1)\tau r\tau s} - 2\, {\hat {\bar
 \epsilon}}_{(-)(r)}\, {\hat {\bar n}}_{(u)}\, {\hat {\bar
 \epsilon}}_{(-)(v)}\, {\bar W}_{(1)\tau ruv} -\nonumber \\
 &-& {\hat {\bar n}}_{(r)}\, {\hat {\bar
 \epsilon}}_{(-)(s)}\, {\hat {\bar n}}_{(u)}\, {\hat {\bar
 \epsilon}}_{(-)(v)}\, {\bar W}_{(1)rsuv}\Big] =\nonumber \\
 &=& {{\sgn}\over 4}\, \Big({\hat {\bar \epsilon}}_{(1)(r)}\, {\hat {\bar \epsilon}}_{(1)(s)}
 - {\hat {\bar \epsilon}}_{(2)(r)}\, {\hat {\bar \epsilon}}_{(2)(s)} -
 i\, ({\hat {\bar \epsilon}}_{(1)(r)}\, {\hat {\bar \epsilon}}_{(2)(s)}
 + {\hat {\bar \epsilon}}_{(2)(r)}\, {\hat {\bar \epsilon}}_{(1)(s)})\Big)\nonumber \\
 &&\Big[\delta_{rs}\, \triangle\, (\Gamma_r^{(1)} + 2\,
 \phi_{(1)}) + \partial_r\, \partial_s\, (2\, \phi_{(1)} -
 \Gamma_r^{(1)} - \Gamma_s^{(1)}) -\nonumber \\
 &-& {\hat {\bar n}}_{(u)}\, \Big(\sum_w\, ((1 - \delta_{ru})\, \delta_{uw}\, \partial_s - (1 - \delta_{rs})\, \delta_{sw}\, \partial_u)\,(\partial_r\, {\bar n}_{(1)(w)} + \partial_w\, {\bar n}_{(1)(r)}) +\nonumber \\
 &+& 2\, (\delta_{ru}\, \partial_s - \delta_{rs}\, \partial_u)\,
 ({1\over 3}\, {}^3K_{(1)} - {{8 \pi G}\over {c^3}}\, \sum_{\bar a}\, \gamma_{\bar ar}\, \Pi_{\bar a}
  + \partial_r\, {\bar n}_{(1)(r)} -
 {1\over 3}\, \sum_c\, \partial_c\, {\bar n}_{(1)(c)}) \Big) -\nonumber \\
 &-& {\hat {\bar n}}_{(u)}\, {\hat {\bar n}}_{(v)}\, \Big(
 (\delta_{sv}\, \partial_u - \delta_{uv}\, \partial_s)\, \partial_r\,
 (\Gamma_v^{(1)} + 2\, \phi_{(1)}) +
 (\delta_{ru}\, \partial_s - \delta_{rs}\, \partial_u)\, \partial_v\,
 (\Gamma_r^{(1)} + 2\, \phi_{(1)})  \Big)\Big]  =\nonumber \\
 &=& \Psi_{(1)4(R)}^{(W)} + i\, \Psi_{(1)4(I)}^{(W)}.
 \label{5.2}
 \eea

\medskip

Therefore the linearized Weyl scalars are linear functions of $R_{\bar a}$, $\Pi_{\bar a}$, ${}^3K_{(1)}$ and of their gradients.

\subsection{The Weyl Eigenvalues}

For the Weyl eigenvalues of Eq.(\ref{2.9}) we get

\bigskip

\bea
 w_{(1) 1} + i\, w_{(1) 2} &\cir&  {\bar w}_{(1) 1} + i\, {\bar w}_{(1) 2} =\nonumber \\
  &=& 2\, \Psi^{(W)}_{(1)0}\, \Psi^{(W)}_{(1)4} - 8\, \Psi_1\, \Psi^{(W)}_{(1)3} + 6\,
\Psi^{(W) 2}_{(1)2} = O(\zeta^2),\nonumber \\
 {}&&\nonumber \\
 w_{(1) 3} + i\, w_{(1) 4} &\cir&  {\bar w}_{(1) 3} + i\, {\bar w}_{(1) 4} =\nonumber \\
  &=& 6\, \Big(\Psi^{(W)}_{(1)0}\, \Psi^{(W)}_{(1)2}\, \Psi^{(W)}_{(1)4} + 2\, \Psi^{(W)}_{(1)1}\,
\Psi^{(W)}_{(1)2}\, \Psi^{(W)}_{(1)3} - \Psi^{(W) 3}_{(1)2} -\nonumber \\
 &-&\Psi^{(W)}_{(1)0}\, \Psi^{(W) 2}_{(1)3} - \Psi^{(W)}_{(1)4}\, \Psi^{(W) 2}_{(1)1}\Big) = O(\zeta^3),\nonumber \\
{}&&
\label{5.3}
\eea

\bigskip

To have quantities of order $O(\zeta)$to be used in Eqs.(\ref{4.5}) we must consider functions of the Weyl eigenvalues like either $|{\bar w}_{(1) 1}|^{1/2}$, $|{\bar w}_{(1) 2}|^{1/2}$,
$|{\bar w}_{(1) 3}|^{1/3}$, $|{\bar w}_{(1) 4}|^{1/3}$ or  ${\bar w}_{(1) 3}/{\bar w}_{(1) 1}$, ${\bar w}_{(1) 3}/{\bar w}_{(1) 2}$, ${\bar w}_{(1) 4}/{\bar w}_{(1) 1}$,
${\bar w}_{(1) 4}/{\bar w}_{(1) 2}$.

\section{Conclusions}

In these two papers we have found the Hamiltonian expression of the 4-Riemann and 4-Weyl tensors in the framework of the York basis of canonical ADM tetrad gravity described by using radar 4-coordinates adapted to a 3+1 splitting of the asymptotically Minkowskian space-time.

\medskip

This Hamiltonian description, till now not present in the literature, identifies Hamiltonian radar tensors, which
coincide with the 4-Riemann and 4-Weyl tensors on-shell on the solutions of Einstein's equations. Due to the use of radar 4-coordinates each component of these Hamiltonian radar tensors is a 4-scalar of the space-time.

\medskip

We have also introduced a set of Hamiltonian null tetrads, which allow us to obtain the Hamiltonian expression of the Weyl scalars and of the Weyl eigenvalues. These null tetrads can be used to make a Hamiltonian reformulation of the whole Newman-Penrose formalism.

\medskip

In this paper we have also discussed the problem of the determination of the DO's of the gravitational field and shown that the use of radar 4-coordinates allows us to define 4-scalar DO's which are also BO's.

\medskip

We have shown that most probably the Weyl eigenvalues are not BO's and that their use is relevant only for the physical identification of the mathematical points of the space-time 4-manifold as point-events labeled by the gravitational field. Therefore, the gravitational field does not only describe the metric structure of the mathematical space-time, but also gives a physical reality to it, differently from what happens with Minkowski and Galilei space-times which are mathematical 4-manifolds absolutely given.

\bigskip

The main open problem is now the determination of solutions of the super-Hamiltonian and super-momentum constraints. As we have shown this would allow an explicit determination of the DO's of the gravitational field by means of of Shanmugadhasan canonical transformation to a canonical basis adapted to all the first class constraints.

\medskip

A connected open problem is to try to understand which are the implications of our Hamiltonian description of the gravitational field for the canonical quantization of gravity. Is only the metric structure of the space-time to be quantized or also the space-time 4-manifold has to be replaced with a more complex (quantized? non-commutative?...) structure?

\appendix

\section{The Bel-Robinson Tensor and the Second Order Invariants of the 4-Riemann and 4-Weyl Tensors}

The results of paper I and Section II allows us to find the Hamiltonian expression of the Bel-Robinson and of the second order invariants of the 4-Riemann and 4-Weyl tensors.

\subsection{The Bel-Robinson Tensor}

By using the expression of the Bel-Robinson tensor ${}^4T_{ABCD} = {}^4T_{(ABCD)}$ (${}^4T^A{}_{ACD} = 0$ )
in terms of the 4-Weyl tensor of Ref. \cite{18}  ($\eta_{ABCD} = \sgn\, \epsilon_{ABCD} / (1 + n)\, \sqrt{{}^3g}$,
$\epsilon_{\tau 123} = 1$) we can get its on-shell Hamiltonian expression
(see Eq.I-(2.10) for the 4-metric and (\ref{2.3}) for the 4-Weyl tensor; $(..)$ and $[..]$ mean symmetrization and anti- symmetrization respectively)

\bea
 {}^4T_{ABCD} &=& {}^4C_{AECF}\, {}^4C_B{}^E{}_D{}^F + {1\over 4}\,
 \eta_{AE}{}^{HI}\, \eta_B{}^{EJ}{}_K\, {}^4C_{HICF}\,
 {}^4C_J{}^K{}_D{}^F =\nonumber \\
 &&{}\nonumber \\
 &=&{}^4C_{AECF}\, {}^4C_B{}^E{}_D{}^F - {3\over 2}\,
 {}^4g_{A[B}\, {}^4C_{JK]CF}\, {}^4C^{JK}{}_D{}^F \cir \nonumber \\
 &\cir& {\bar {\cal C}}_{AECF}\, {\bar {\cal C}}_B{}^E{}_D{}^F - {3\over 2}\,
 {}^4g_{A[B}\, {\bar {\cal C}}_{JK]CF}\, {\bar {\cal C}}^{JK}{}_D{}^F.
 \label{a1}
 \eea

 \medskip

 The Bianchi identities for the Riemann tensor imply that in
 vacuum we have $\nabla^A\, {}^4T_{ABCD} \cir
 0$.

\subsection{The Second Order Invariants of the 4-Riemann and 4-Weyl Tensors.}

As shown in Refs. \cite{19,20} the 4-Riemann tensor
and the 4-Weyl tensor have three ($k_1, k_2, k_3$) and two ($I_1, I_2$) second order 4-scalar invariants
respectively. By using Eqs.(\ref{2.3}) and Einstein's equations I-(2.16) we can get their on-shell Hamiltonian
expressions.

\bigskip

1) For the {\it Kretschmann invariant} $k_1$  and for $I_1$ we have

\bea
  I_1 &=& {}^4C_{ABCD}\, {}^4C^{ABCD} \cir {\bar {\cal C}}_{ABCD}\, {\bar {\cal C}}^{ABCD},\nonumber \\
 {}&&\nonumber \\
 k_1 &=& {}^4R_{ABCD}\, {}^4R^{ABCD} = I_1 + 2\, {}^4R_{AB}\,
 {}^4R^{AB} - {1\over 3}\, {}^4R^2 \cir\nonumber \\
 &\cir& I_1 + 2\, ({{8 \pi G}\over {c^3}})^2 + {\hat T}_{AB}\, {\hat T}^{AB}.
 \label{a2}
 \eea

\bigskip

2) For $I_2$ and the {\it Chern-Pontryagin invariant} $k_2$ we have
($[{}^*R]_{ABCD} = {1\over 2}\, \eta_{ABEF}\,
R^{EF}{}_{CD}$ is the dual of the 4-Riemann tensor;
the double dual is $[{}^*R^*]_{ABCD} = {1\over
4}\, \eta^{ABMN}\, R_{MN}{}^{EF}\, \eta_{EFCD}$; for the 4-Weyl tensor we have the
same definition and moreover the self-duality
property $[{}^*C^*] = - C$)

\beq
 k_2 = [{}^*{}^4R]_{ABCD}\, {}^4R^{ABCD} = [{}^*{}^4C]_{ABCD}\, {}^4C^{ABCD} = I_2
 \cir [{}^*{\bar {\cal C}}]_{ABCD}\, {\bar {\cal C}}^{ABCD}.
 \label{a3}
 \eeq

 \bigskip

3) Finally the {\it Euler invariant} is $k_3 = [{}^*{}^4R^*]_{ABCD}\, {}^4R^{ABCD} = - I_1 + 2\, {}^4R_{AB}\, {}^4R^{AB} - {2\over 3}\, {}^4R^2$.

\section{The 3-Orthogonal Schwinger Time Gauges and their Linearization}

In this Appendix we review the family of 3-orthogonal Schwinger time gauges and the HPM linearization of canonical ADM tetrad gravity defined in Refs.\cite{6,7}. We give only the results in absence of matter to simplify the discussion. Then we discuss what can be said about the HPM linearization in gauges different (but near) from the 3-orthogonal ones.

\subsection{The 3-Orthogonal Schwinger Time Gauges}

This family of non-harmonic gauges is chosen in such a way that the 3-metric in the 3-spaces is
diagonal and is defined by the gauge fixings

\bea
 &&\varphi_{(a)}(\tau, \sigma^u) \approx 0,\qquad \alpha_{(a)}(\tau, \sigma^u) \approx
 0,\nonumber \\
 &&{}\nonumber \\
 &&\theta^i(\tau, \sigma^u) \approx 0,\qquad {}^3K(\tau, \sigma^u) = {{12 \pi G}\over {c^3}}\,
 \pi_{\tilde \phi}(\tau, \sigma^u) \approx F(\tau, \sigma^u),
 \label{b1}
 \eea

\noindent This family, parametrized by the arbitrary function $F(\tau, \sigma^u)$,
implies $ 0 = \partial_{\tau}\, \varphi_{(a)}(\tau, \sigma^u) \cir \lambda_{\varphi_{(a)}}(\tau, \sigma^u)$ and
$0 = \partial_{\tau}\, \alpha_{(a)}(\tau, \sigma^u) \cir \lambda_{\alpha_{(a)}}(\tau, \sigma^u)$.
The $\tau$-preservation of the second half of  Eqs.(\ref{b1}) (the kinematical Hamilton equations $0 = \partial_{\tau}\, \theta^i(\tau, \sigma^u) \cir ...$ and the dynamical Raychaudhuri equation $\partial_{\tau}\, F(\tau, \sigma^u) \cir
\partial_{\tau}\, {}^3K(\tau, \sigma^u) \cir ...$ given in Eqs.(2.12) and (4.5) of Ref. \cite{5} respectively) generates four coupled elliptic PDE in the 3-spaces for the lapse and shift functions
\footnote{Instead in the 4-harmonic gauges the lapse and shift functions obey hyperbolic PDE as shown in Eq. (5.4) of Ref. \cite{5}.}.

\medskip

In Refs.\cite{5,6} there is the explicit form of the Hamilton
equations (with matter included) for all the canonical variables of the York basis
in the Schwinger time gauges and their restriction to the 3-orthogonal ones.

\medskip

In the family of 3-orthogonal gauges the
super-Hamiltonian and super-momentum constraints  are coupled
{\it elliptic} PDE for their unknowns, whose  expression is given
in Ref.\cite{5}. In these gauges the solutions $\tilde \phi$ and $\pi_i^{(\theta)}$
(or $\sigma_{(a)(b)}{|}_{a \not= b}$) of the super-Hamiltonian and
super-momentum constraints on the 3-spaces $\Sigma_{\tau}$ are
functionals of the tidal variables $R_{\bar a}$, $\Pi_{\bar a}$, of
the York time ${}^3K \approx F$ and of the matter, all evaluated on
$\Sigma_{\tau}$.

\bigskip

Therefore, given  the Cauchy data for the tidal variables on an
initial 3-space, one can find a solution of of their hyperbolic PDE, which then can be rewritten as a solution of
Einstein's equations in radar 4-coordinates adapted to a time-like observer in the chosen
gauge.

\medskip

Therefore, as shown in Ref.\cite{5}, in these gauges only the tidal variables (the
gravitational waves after linearization), and therefore only the
eigenvalues of the 3-metric with unit determinant inside
$\Sigma_{\tau}$, depend (in a retarded way) on the no-incoming
radiation condition.

\bigskip

In these gauges the Hamiltonian variables of the York basis, the 3-geometry of the 3-spaces (3-Christoffel
symbols and 3-Riemann tensor) and the 4-Christoffel symbols are given by Eqs. I-(2.10), I-(2.12), I-(A1) and I-(3.2) by
putting $R_{(a)(b)}(\alpha_{(e)} = 0) = \delta_{(a)(b)}$ and $V_{ra}(\theta^i = 0) = \delta_{ra}$.

\bigskip

In the family of 3-orthogonal Schwinger time gauges the York canonical basis in absence of matter becomes

\bea
&&\begin{minipage}[t]{4 cm}
\begin{tabular}{|ll|ll|l|l|l|} \hline
$\varphi_{(a)} \approx 0$ & $\alpha_{(a)} \approx 0$ & $n$ & ${\bar n}_{(a)}$ &
$\theta^r \approx 0$ & $\tilde \phi$ & $R_{\bar a}$\\ \hline
$\pi_{\varphi_{(a)}} \approx0$ &
 $\pi^{(\alpha)}_{(a)} \approx 0$ & $\pi_n \approx 0$ & $\pi_{{\bar n}_{(a)}} \approx 0$
& $\pi^{(\theta )}_r$ & $\pi_{\tilde \phi} \approx {{c^3}\over {12 \pi G}}\, F$ & $\Pi_{\bar a}$ \\
\hline
\end{tabular}
\end{minipage}.\nonumber \\
{}&&
 \label{b2}
 \eea

To it one has to add the constraints $n(\tau, \sigma^u) - n^{(S)}(\tau, \sigma^u) \approx 0$ and ${\bar n}_{(a)}(\tau,
\sigma^u) - {\bar n}_{(a)}^{(S)}(\tau, \sigma^u) \approx 0$, where $n^{(S)}(\tau, \sigma^u)$ and ${\bar n}_{(a)}^{(S)}(\tau, \sigma^u)$ are the solutions for the lapse and shift functions implied by the $\tau$-constancy of the gauge fixings. Since there are 10 pairs of second class constraints ($\varphi_{(a)}(\tau, \sigma^u) \approx 0$ and $\pi_{\varphi_{(a)}}(\tau, \sigma^u) \approx 0$, $\alpha_{(a)}(\tau, \sigma^u) \approx 0$ and $\pi_{\alpha_{(a)}} \approx 0$, $n(\tau, \sigma^u) - n^{(S)}(\tau, \sigma^u) \approx 0$ and $\pi_n(\tau, \sigma^u) \approx 0$, ${\bar n}_{(a)}(\tau, \sigma^u) - {\bar n}_{(a)}^{(S)}(\tau, \sigma^u) \approx 0$ and $\pi_{{\bar n}_{(a)}} \approx 0$), one has to evaluate the Dirac brackets and then to find the new canonical basis $({\check \theta}^i, {\check \pi}_i^{(\theta)}, {\check {\tilde \phi}}, {\check \pi}_{\tilde \phi}, {\check R}_{\bar a}, {\check \Pi}_{\bar a})$ of the reduced phase space. In this new basis one has to find the form of the remaining 4 pairs of second class constraints (the super-Hamiltonian and super-momentum constraints and their gauge fixings $\pi_{\tilde \phi} \approx {{c^3}\over {12 \pi G}}\, F$ and $\theta^i \approx 0$), find again the Dirac brackets and the final physical tidal variables.

\subsection{The Linearized Theory in Absence of Matter in the
3-Orthogonal Schwinger Time Gauges}

As shown in Ref.\cite{6} the HPM linearization in  the 3-orthogonal
Schwinger time gauges (after the addition of a suitable ultra-violet cutoff on the matter, when present) implies ($\zeta$ is the first order quantity defining the weak field approximation)

\begin{eqnarray*}
 &&R_{\bar a} = \sum_a \gamma_{\bar aa}\, \Gamma_a^{(1)} = O(\zeta),
 \nonumber \\
 &&\tilde \phi = 1 + 6\, \phi_{(1)},\qquad N = 1 + n_{(1)},\qquad
 {\bar n}_{(a)} = {\bar n}_{(1)(a)},\qquad ({\tilde \phi}^{-1}\,
 \partial_r\, \tilde \phi = 6\, \partial_r\, \phi_{(1)}),\nonumber \\
 &&{}^3K_{(1)} = {{12\pi G}\over {c^3}}\, \pi_{(1)\tilde \phi} =
 {1\over L}\, O(\zeta) \approx F_{(1)},\nonumber \\
 &&{}\nonumber \\
  &&{}^3{\bar e}_{(1)(a)r} = (1 + \Gamma_r^{(1)} + 2\, \phi_{(1)})\,
  \delta_{ar},\nonumber \\
 &&{}^3{\bar e}^r_{(1)(a)} = (1 - \Gamma_r^{(1)} - 2\, \phi_{(1)})\,
 \delta_{ar}, \nonumber \\
 &&{}\nonumber \\
 &&{}^4g_{(1)\tau\tau} = \sgn\, (1 + 2\, n_{(1)}),\qquad
 {}^4g_{(1)\tau r} = - \sgn\, {\bar n}_{(1)(r)},\nonumber \\
 &&{}^4g_{(1)rs} = - \sgn\, {}^3g_{(1)rs} = - \sgn\, [1 + 2\,
 (\Gamma_r^{(1)} + 2\, \phi_{(1)})]\, \delta_{rs},
 \end{eqnarray*}

\bea
 &&{{8\pi G}\over {c^3}}\, \Pi_{\bar a} \cir \partial_{\tau}\,
 R_{\bar a} - \sum_a\, \gamma_{\bar aa}\, \partial_a\, {\bar
 n}_{(1)(a)} = {1\over L}\, O(\zeta),\nonumber \\
 &&{}\nonumber \\
 &&{}^3K_{(1)rs} = (1 - \delta_{rs})\, \sigma_{(1)(r)(s)} +
 \delta_{rs}\, [{1\over 3}\, {}^3K_{(1)} + \sigma_{(1)(r)(r)} +
 \nonumber \\
 &&+ \partial_r\, {\bar n}_{(1)(r)} - {1\over 3}\, \sum_c\,
 \partial_c\, {\bar n}_{(1)(c)}] = {1\over L}\, O(\zeta).
 \label{b3}
 \eea

\bigskip

For the 3-geometry we get from Eqs. I-(2.12) and I-(A1) ($\triangle = \sum_a\, \partial_a^2$)

\bea
 {}^3\Gamma^r_{uv} &=& {1\over 3}\, (\delta_{ru}\, \partial_v\,
 \phi_{(1)} + \delta_{rv}\, \partial_u\, \phi_{(1)} - \delta_{uv}\,
 \partial_r\, \phi_{(1)}) +\nonumber \\
 &+& \sum_{\bar a}\, \gamma_{\bar ar}\, (\delta_{ur}\,
 \partial_v\, R_{\bar a} + \delta_{vr}\, \partial_u\, R_{\bar a} -
 \delta_{uv}\, \partial_r\, R_{\bar a}) +\nonumber \\
 &&{}\nonumber \\
 {}^3R_{rsuv} &=& \delta_{sv}\, \partial_r\, \partial_u\,
 (\Gamma_s^{(1)} + 2\, \phi_{(1)}) - \delta_{rv}\, \partial_s\,
 \partial_u\, (\Gamma_r^{(1)} + 2\,  \phi_{(1)}) +\nonumber \\
 &+& \delta_{ru}\, \partial_s\,
 \partial_v\, (\Gamma_r^{(1)} + 2\, \phi_{(1)}) - \delta_{su}\,
 \partial_r\, \partial_v\, (\Gamma_s^{(1)} + 2\,
 \phi_{(1)}),\nonumber \\
 &&{}\nonumber \\
 {}^3R_{sv} &=& \delta_{sv}\, \triangle\, (\Gamma_s^{(1)} + 2\,
 \phi_{(1)}) + \partial_s\, \partial_v\, (2\, \phi_{(1)} -
 \Gamma_s^{(1)} - \Gamma_v^{(1)}),\nonumber \\
 &&{}\nonumber \\
 {}^3R &=& 2(4\, \triangle\, \phi_{(1)} - \sum_s\, \partial_s^2\,
 \Gamma_s^{(1)}).
 \label{b4}
 \eea

\bigskip

In absence of matter \footnote{In Ref. \cite{6} there is the solution for the case in which the matter consists of charged positive-energy scalar particles plus the electro-magnetic field.}
the solution of the super-Hamiltonian and super-momentum constraints is

\bea
 \phi_{(1)} &\approx& - {1\over 4}\, \sum_c\, {{\partial_c^2}\over
 {\triangle}}\, \Gamma_c^{(1)},\nonumber \\
 &&{}\nonumber \\
 \sigma_{(1)(a)(b)}{|}_{a \not= b} &\approx& {1\over 2}\, (\partial_a\,
 {\bar n}_{(1)(b)} + \partial_b\, {\bar n}_{(1)(a)}),\nonumber \\
 &&{}\nonumber \\
 &&\Downarrow\nonumber \\
 &&{}\nonumber \\
 {{8\pi G}\over {c^3}}\, \pi_i^{(\theta)}  &=& \sum_{a \not= b}\,
 \epsilon_{iab}\, (\Gamma_a^{(1)} - \Gamma_b^{(1)})\, \sigma_{(1)(a)(b)}
 = {1\over L}\, O(\zeta^2)\approx 0.
 \label{b5}
 \eea

\bigskip

The time constancy of the gauge fixings $\theta^i \approx 0$ and
${}^3K_{(1)} \approx F_{(1)}$ are coupled elliptic equations for the
lapse and shift functions. They are the kinematical Hamilton equations
 for $0 \approx \partial_{\tau}\, \theta^i \cir ...$ and
the dynamical Raychaudhuri equation for $\partial_{\tau}\, F_{(1)}
\approx \partial_{\tau}\, {}^3K_{(1)} \cir ...$, whose linearized solution is
\medskip

\bea
 n_{(1)} &\approx& n_{(1)}^{(S)} =
 - {{\partial_{\tau}}\over {\triangle}}\, {}^3K_{(1)} =
 - {{12 \pi G}\over {c^3}}\, {{\partial_{\tau}}\over {\triangle}}\, \pi_{(1) \tilde \phi},\nonumber \\
 {\bar n}_{(1)(a)} &\approx& {{\partial_a}\over {\triangle}}\, {}^3K_{(1)} + {1\over 2}\,
 {{\partial_a}\over {\triangle}}\, \partial_{\tau}\, \Big(4\,
 \Gamma_a^{(1)} - \sum_c\, {{\partial_c^2}\over {\triangle}}\,
 \Gamma_c^{(1)}\Big).
 \label{b6}
 \eea

\medskip

To express the solution (\ref{b6}) for ${\bar n}_{(1)(a)}$ as a Hamiltonian constraint
we must use the equation  $\partial_{\tau}\, \Gamma_a^{(1)} \cir - {{8 \pi G}\over {c^3}}\,
\sum_{\bar a}\, \gamma_{\bar aa}\, \Pi_{\bar a} + \partial_a\, {\bar n}_{(1)(a)} - {1\over 3}\,
\sum_c\, \partial_c\, {\bar n}_{(1)(c)}$ (see after I-(2.11)) to eliminate the velocity
$\partial_{\tau}\, \Gamma_a^{(1)}$. Its use gives (${\hat Z}_{ab}$ is an elliptic operator in the 3-space and
${\hat Z}^{-1}_{ab}$ is its inverse)

\begin{eqnarray*}
 &&\Big(1 - 2\, {{\partial_a}\over {\triangle}}\Big)\, {\bar
 n}_{(1)(a)} + {1\over 2}\, {{\partial_a}\over {\triangle}}\,
 \sum_c\, \Big(1 + {{\partial_c^2}\over {\triangle}}\Big)\,
 \partial_c\, {\bar n}_{(1)(c)} \approx\nonumber \\
 &&\approx {{\partial_a}\over {\triangle}}\, {}^3K_{(1)} + {{4 \pi G}\over {c^3}}\,
  {{\partial_a}\over {\triangle}}\, \sum_b\,
  \Big(4\, \delta_{(a)(b)} -  {{\partial_b^2}\over
 {\triangle}}\Big)\, \sum_{\bar a}\, \gamma_{\bar ab}\, \Pi_{\bar a},\nonumber \\
 \Downarrow&&
\end{eqnarray*}

 \begin{eqnarray*}
 \sum_b\, {\hat Z}_{ab}\, {\bar n}_{(1)(b)} &{\buildrel {def}\over =}& \sum_b\,
 \Big[(1 - 2\, {{\partial_a^2}\over {\triangle}})\, \delta_{ab} +
  {{\partial_a\, \partial_b}\over {\triangle}}\Big]\, {\bar n}_{(1)(b)} \approx\nonumber \\
 &\approx& {4\over 3}\, {{\partial_a^2}\over {\triangle}}\, {}^3K_{(1)} + {{16 \pi G}\over {c^3}}\,
 {{\partial_a^2}\over {\triangle}}\, \sum_{\bar a}\, \gamma_{\bar aa}\, \Pi_{\bar a},\nonumber \\
 \Downarrow &&
 \end{eqnarray*}

 \bea
 {\bar n}_{(1)(a)} &\approx& {\bar n}_{(1)(a)}^{(S)} = {{16 \pi G}\over {c^3}}\,
 \sum_b\, {\hat Z}^{-1}_{ab}\, {{\partial^2_b}\over {\triangle}}\,
 (\pi_{(1) \tilde \phi} + \sum_{\bar a}\, \gamma_{\bar ab}\, \Pi_{\bar a}).\nonumber \\
 {}&&
 \label{b7}
 \eea

Eq.(\ref{b7}) gives the Hamiltonian expression of the linearized shift function of the 3-orthogonal
gauges.

\bigskip

Once the lapse and shift functions have been determined, we have the
expressions $\lambda_n \cir \partial_{\tau}\, n_{(1)}$ and
$\lambda_{{\bar n}_{(1)(a)}} \cir \partial_{\tau}\, {\bar n}_{(1)(a)}$
for the remaining Dirac multipliers.

\subsection{The Linearization in Arbitrary Gauges near the 3-Orthogonal Gauges}

Let us now consider the possibility of studying the HPM linearized theory by relaxing the gauge fixings (\ref{b1}) with
the only restriction to be near the 3-orthogonal gauge where the 3-metric is diagonal. Therefore we require
$\theta^i(\tau, \sigma^u) = \theta^i_{(1)}(\tau, \sigma^u) = O(\zeta) \not= 0$, so that we have $V_{ra}(\theta^i_{(1)}) = \delta_{ra} - \epsilon_{rai}\, \theta^i_{(1)}$ (see after Eq. I-(2.9)), and we put no restriction on the York time ${}^3K_{(1)}(\tau, \sigma^u) = {{12 \pi G}\over {c^3}}\, \pi_{\tilde \phi (1)}(\tau, \sigma^u)$. The relax of the conditions $\varphi_{(a)}(\tau, \sigma^u) \approx 0$ and $\alpha_{(a)}(\tau, \sigma^u) \approx 0$ defining the Schwinger time gauges is irrelevant for all the quantities depending only on the 4-metric. Only the general cotetrads I-(2.5) depend upon these six gauge variables.

\medskip

In Eqs.(\ref{b3}) there is the following modification of the triads and cotriads
${}^3{\bar e}_{(1)(a)r} = (1 + \Gamma_r^{(1)} + 2\, \phi_{(1)})\,
\delta_{ar} - \epsilon_{rai}\, \theta^i_{(1)}$,
${}^3{\bar e}^r_{(1)(a)} = (1 - \Gamma_r^{(1)} - 2\, \phi_{(1)})\,
\delta_{ar} - \epsilon_{rai}\, \theta^i_{(1)}$, which leaves the linearized 3-metric in the
diagonal form (the deviations from 3-orthogonality appear at the second HPM order).
As can be checked from Eqs. I-(2.12) and I-(A1) the 3-geometry on the 3-space (namely the 3-Christoffel symbols
and the 3-Riemann and 3-Ricci tensors) do not depend on $\theta^i_{(1)}$.
\medskip

The solution of the super-Hamiltonian and super-momentum constraints
does not depend on $\theta^i_{(1)}$ .

\medskip

For the super-Hamiltonian constraint we have the solution $\tilde
\phi \approx 1 + 6\, \phi_{(1)}$ with $\phi_{(1)}$ of Eq.(\ref{b5}) also for $\theta^i(\tau, \vec
\sigma) = \theta^i_{(1)}(\tau, \vec \sigma) = O(\zeta) \not= 0$.

The Raychaudhuri equation (4.5) of Ref.\cite{6}, $\partial_{\tau}\,
{}^3K_{(1)} \cir \triangle\, n_{(1)}$, remains
unchanged if $\theta^i(\tau, \vec \sigma) = \theta^i_{(1)}(\tau,
\vec \sigma) = O(\zeta) \not= 0$. Therefore the determination of the
lapse function depends on the $\tau$-preservation of the gauge fixing for
${}^3K_{(1)}$, when added, like in 3-orthogonal gauges.

\medskip

The solution of the super-momentum constraints with $\theta^i(\tau,
\vec \sigma) = \theta^i_{(1)}(\tau, \vec \sigma) = O(\zeta) \not= 0$
remains $\sigma_{(1)(a)(b)}{|}_{a \not= b} \approx {1\over 2}\,
\Big(\partial_a\, {\bar n}_{(1)(b)} + \partial_b\, {\bar
n}_{(1)(a)}\Big)$, implying $\pi_i^{(\theta)} \approx 0$ in the York
canonical basis. This result derives from Eqs. (4.5) and (6.9) of Ref.\cite{5}:
if $\partial_{\tau}\, \theta^i_{(1)} \not= 0$ replaces zero in
Eqs.(6.9) of Ref.\cite{5}, then Eq.(6.10) of Ref.\cite{5} (and (2.11) of Ref.\cite{6}) has the extra term
$(Q_a\, Q_b^{-1} - Q_b\, Q_a^{-1})\, \sum_i\, \epsilon_{abi}\,
\partial_{\tau}\, \theta^i_{(1)} = 2\, (\Gamma_a^{(1)} -
\Gamma_b^{(1)})\, \sum_i\, \epsilon_{abi}\,
\partial_{\tau}\, \theta^i_{(1)} + O(\zeta^3) = O(\zeta^2)$ at the
second member. But it is higher order, so that Eqs.(4.8) of Ref.\cite{6} (the
super-momentum constraints as equations for
$\sigma_{(1)(a)(b)}{|}_{a \not= b}$) and (4.9) of Ref.\cite{6} (for the shift
function ${\bar n}_{(1)(a)}$) remain valid also with $\theta^i(\tau,
\vec \sigma) = \theta^i_{(1)}(\tau, \vec \sigma) = O(\zeta) \not=
0$, i.e. near the exact 3-orthogonal gauge $\theta^i \approx 0$.
Therefore Eqs.(\ref{b5}) remain valid.

As a consequence the shift functions are still determined and have the expression implied by Eqs.(\ref{b6}) and (\ref{b7}) like in 3-orthogonal gauges.
\medskip

In conclusion all the gauges of the HPM linearization with $\theta^i(\tau,
\vec \sigma) = \theta^i_{(1)}(\tau, \vec \sigma) = O(\zeta) \not= 0$ are 3-orthogonal
like those with $\theta^i \approx 0$: the deviations from 3-orthogonality appear only at the second HPM order
(the 3-orthogonal gauges are a stationarity point into the space of gauges).
\bigskip

The York canonical basis of the HPM-linearized theory in these gauges takes the form

\bea
  &&\begin{minipage}[t]{3cm}
\begin{tabular}{|ll|ll|l|l|l|} \hline
$\varphi_{(a)}$ & $\alpha_{(a)}$ & $1 + n_{(1)}$ & ${\bar
n}_{(1)(a)}$ & $\theta_{(1)}^r$ & $\tilde \phi \approx 1 + 6\, \phi_{(1)}$ & $R_{\bar a}$ \\
\hline $\pi_{\varphi_{(a)}} \approx 0$ &
 $\pi^{(\alpha)}_{(a)} \approx 0$ & $\pi_n \approx 0$ & $\pi_{{\bar n}_{(a)}} \approx 0$
& $\pi^{(\theta )}_r$ & $\pi_{\tilde \phi}$ & $\Pi_{\bar a}$ \\
\hline
\end{tabular}
 \end{minipage} \hspace{1cm}\nonumber \\
 &&{}
 \label{b8}
 \eea

To it one has to add the 3 pairs of second class constraints
${\bar n}_{(a)}(\tau, \sigma^u) - {\bar n}_{(a)}^{(S)}(\tau, \sigma^u) \approx 0$ and $\pi_{{\bar n}_{(a)}} \approx 0$
for the evaluation of the Dirac brackets and the determination of the canonical basis of the reduced phase space.

\bigskip

A final Shanmugadhasan canonical transformation like the one of Eqs.(\ref{4.2}) now gives the following final canonical basis

\begin{eqnarray*}
 &&\begin{minipage}[t]{4 cm}
\begin{tabular}{|ll|ll|l|l|l|} \hline
$\varphi_{(a)}$ & $\alpha_{(a)}$ & $1 + n_{(1)}$ & ${\bar
n}_{(1)(a)}$ & $\theta_{(1)}^r$ & $\hat \phi \approx 0$ & ${\hat R}_{\bar a}$ \\
\hline $\pi_{\varphi_{(a)}} \approx 0$ &
 $\pi^{(\alpha)}_{(a)} \approx 0$ & $\pi_n \approx 0$ & $\pi_{{\bar n}_{(a)}} \approx 0$
& $\pi^{(\theta )}_r \approx 0$ & $\pi_{\tilde \phi}$ & ${\hat
\Pi}_{\bar a}$  \\
\hline
\end{tabular}
\end{minipage}\nonumber \\
 &&{}
\end{eqnarray*}

\bea
&&\hat \phi = \tilde \phi - 1 - 6\, \phi_{(1)}
 = \tilde \phi - 1 +  {3\over 2}\, \sum_c\,
{{\partial_c^2}\over {\triangle}}\, \Gamma_c^{(1)}(\tau, \vec
\sigma) \approx 0,\nonumber \\
&&{\hat R}_{\bar a} = R_{\bar a}, \nonumber \\
&&{\hat \Pi}_{\bar a} = \Pi_{\bar a} + {3\over 2}\ \sum_a\, \gamma_{\bar
aa}\, {{\partial_a^2}\over {\triangle}}\, \pi_{\tilde \phi},
 \label{b9}
 \eea

\noindent which however has to be restricted with the second class constraints
${\bar n}_{(a)}(\tau, \sigma^u) - {\bar n}_{(a)}^{(S)}(\tau, \sigma^u) \approx 0$ and $\pi_{{\bar n}_{(a)}} \approx 0$.

 \vfill\eject

\end{document}